\documentclass[fleqn,12pt,twoside]{article}
\usepackage{amsmath} 

\usepackage{graphicx}
\usepackage{color}
\usepackage[figuresright]{rotating}

\newif\ifpdf 
  \ifx\pdfoutput\undefined \pdffalse 
     \else 
       \pdfoutput=1 
        \pdftrue 
 \fi
\ifpdf 
    \newcommand{\exten}{jpg}
\else 
    \newcommand{\exten}{eps}
\fi

\newcommand{\dav}[2][{ }]{{\color{black}{[}}#2{\color{black}{]^{{\color{black}{#1}}}_{\rm av}}}}

\newcommand{\figone}{%
\begin{figure}[h]
   \centering
   \includegraphics[width=0.95\textwidth]{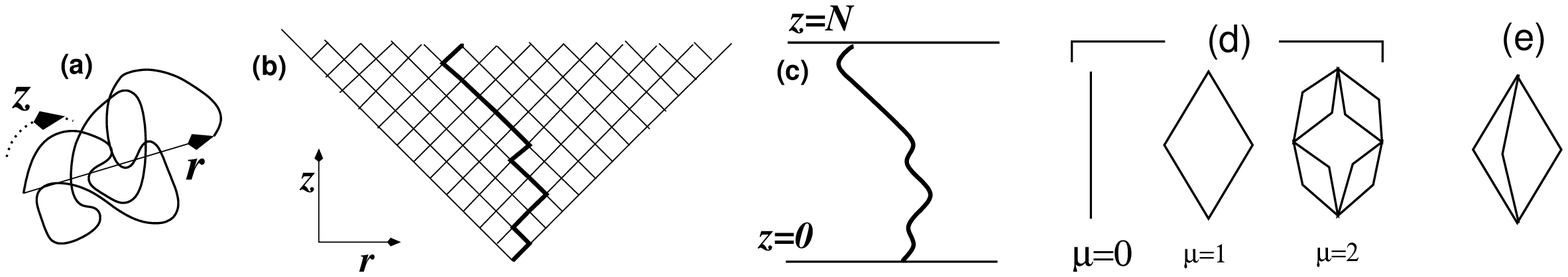}
   \caption{(a) A random walk in $d$ dimensions with $z$ as the
     variable along the contour of the polymer i.e. giving the
     location of the monomers. (b) Directed polymer on a square
     lattice.  A polymer as of (a) can be drawn in $d+1$ dimensions.
     This is like a path of a quantum particle in nonrelativistic
     quantum mechanics. (c) A situation where both the transverse
     space (${\bf r}$) and $z$ are continuous. (d) The directed
     polymers on a hierarchical lattice.  Three generations are
     shown for $4$ bonds. (e) A general motif of $2b$ bonds.} 
   \label{fig:1}
 \end{figure}
}

\newcommand{\figfrc}{%
\begin{figure}[htb]
\begin{minipage}[t]{0.4\textwidth}
    \centering
  \includegraphics[scale=0.9]{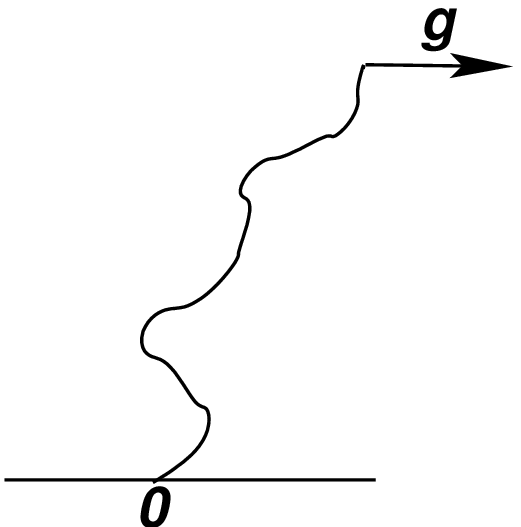}
   \caption{ A directed polymer with an unzipping force. } 
   \label{fig:frc}
\end{minipage}
\hspace{\fill}
\begin{minipage}[t]{0.55\textwidth}
\centering
   \includegraphics[width=1.5in]{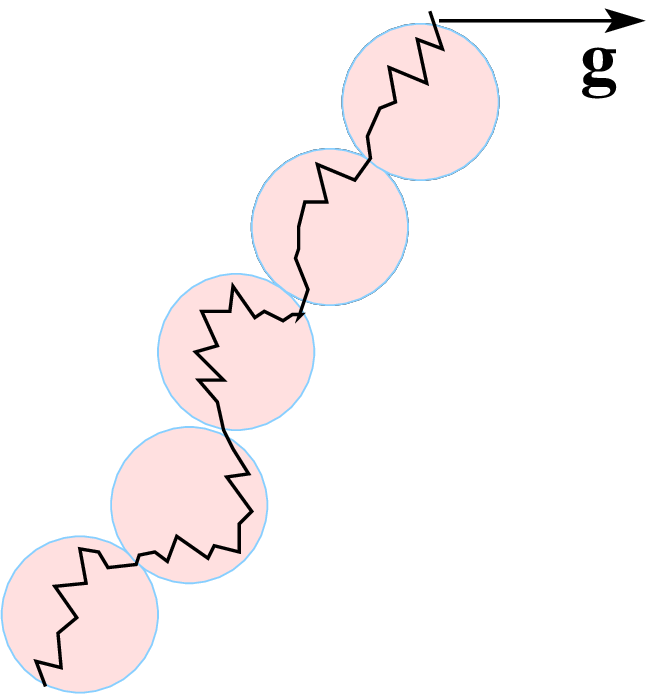}
   \caption{ A blob picture of the polymer under a force.  Though
     drawn as sphere, the $z$-direction is elongated with isotropy in
     the transverse direction.} 
   \label{fig:blob}
\end{minipage}
 \end{figure}
}

\newcommand{\fignN}{%
\begin{figure}[htb]
   \centering
\begin{minipage}[t]{0.55\textwidth}
   \includegraphics[scale=0.9]{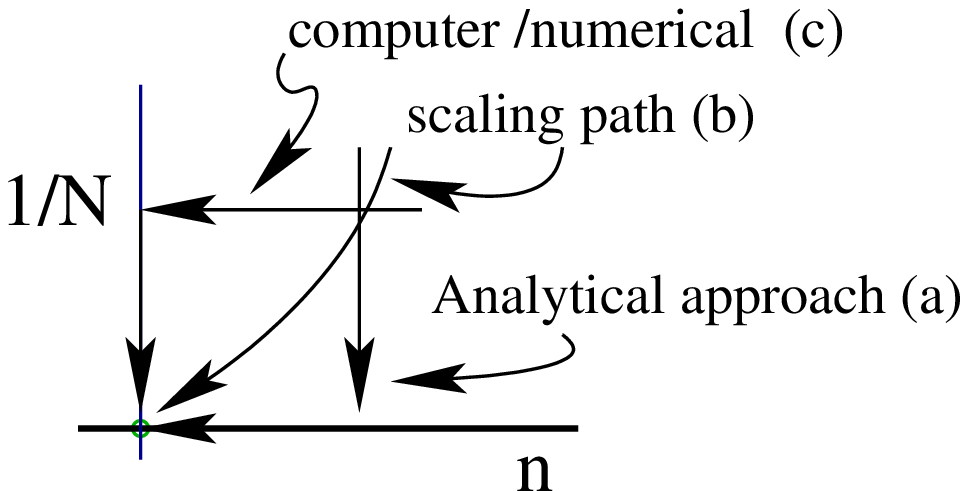}
   \caption{ Paths for replica approach } 
   \label{fig:nN}
\end{minipage}
 \end{figure}
}

\newcommand{\figreu}{%
\begin{figure}[htbp]
   \centering
   \includegraphics[height=1in]{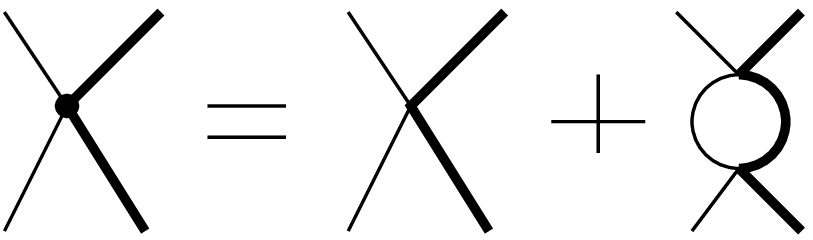}
   \caption{ Renormalization of the interaction. The two polymers are
     represented by the two lines of different thickness and an intersection represents an
     interaction.  A heavy circle on the left hand side represents the
   effective interaction that  is to be used for renormalization.}
   \label{fig:ufp}
 \end{figure}
}

\newcommand{\figfp}{%
\begin{figure}[htbp]
   \centering
   \includegraphics[width=0.5\textwidth]{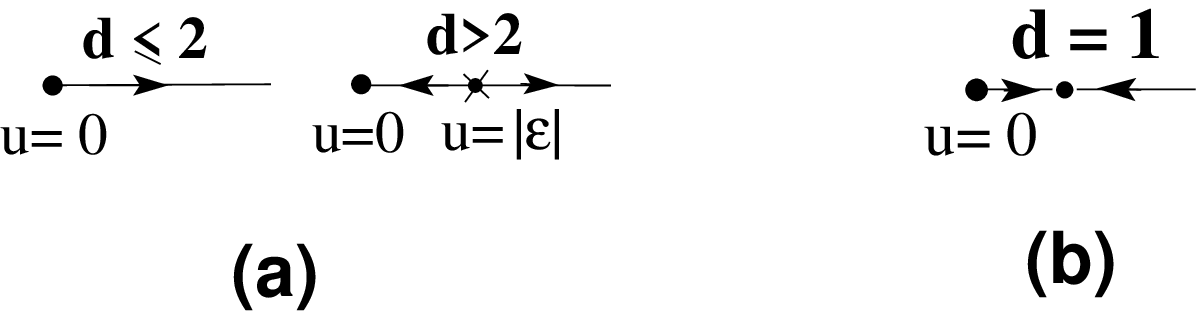}
   \caption{ RG Fixed points for $u$. (a) Based on the second moment
     of the partition function. (b) Based on the KPZ equation.
     Arrows show the flow of $u$. $\epsilon=2-d$. General flow
     diagrams are discussed in App. \ref{app:flwec}.} 
   \label{fig:fp}
 \end{figure}
}

\newcommand{\figriv}{%
\begin{figure}[htbp]
   \centering
   \includegraphics[scale=0.5]{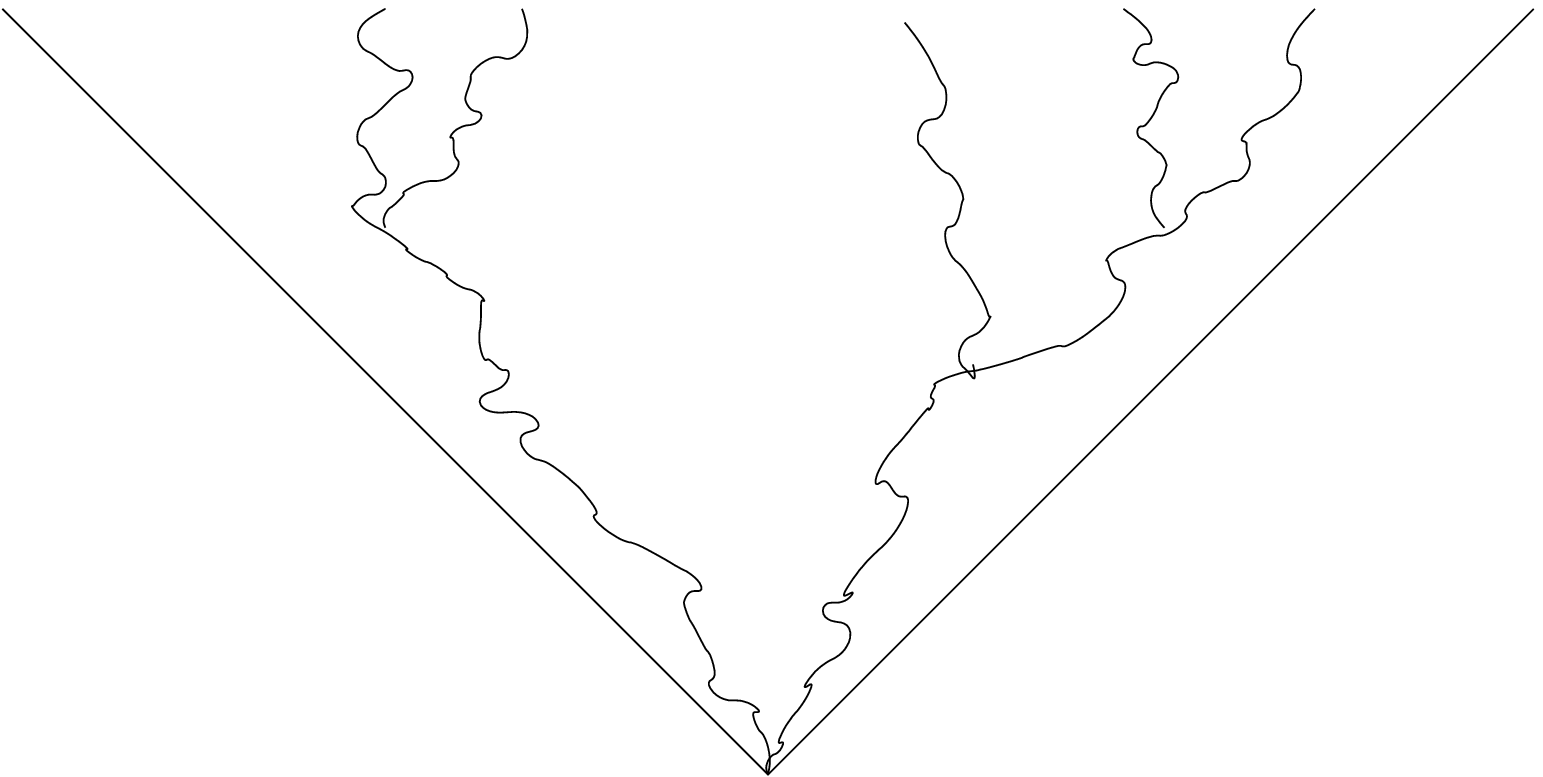}
   \caption{ Various paths for various locations of the end point.
                   } 
   \label{fig:river}
\end{figure}
}

\newcommand{\figuz}{%
\begin{figure}[htbp]
  \centering
   \includegraphics[scale=0.5]{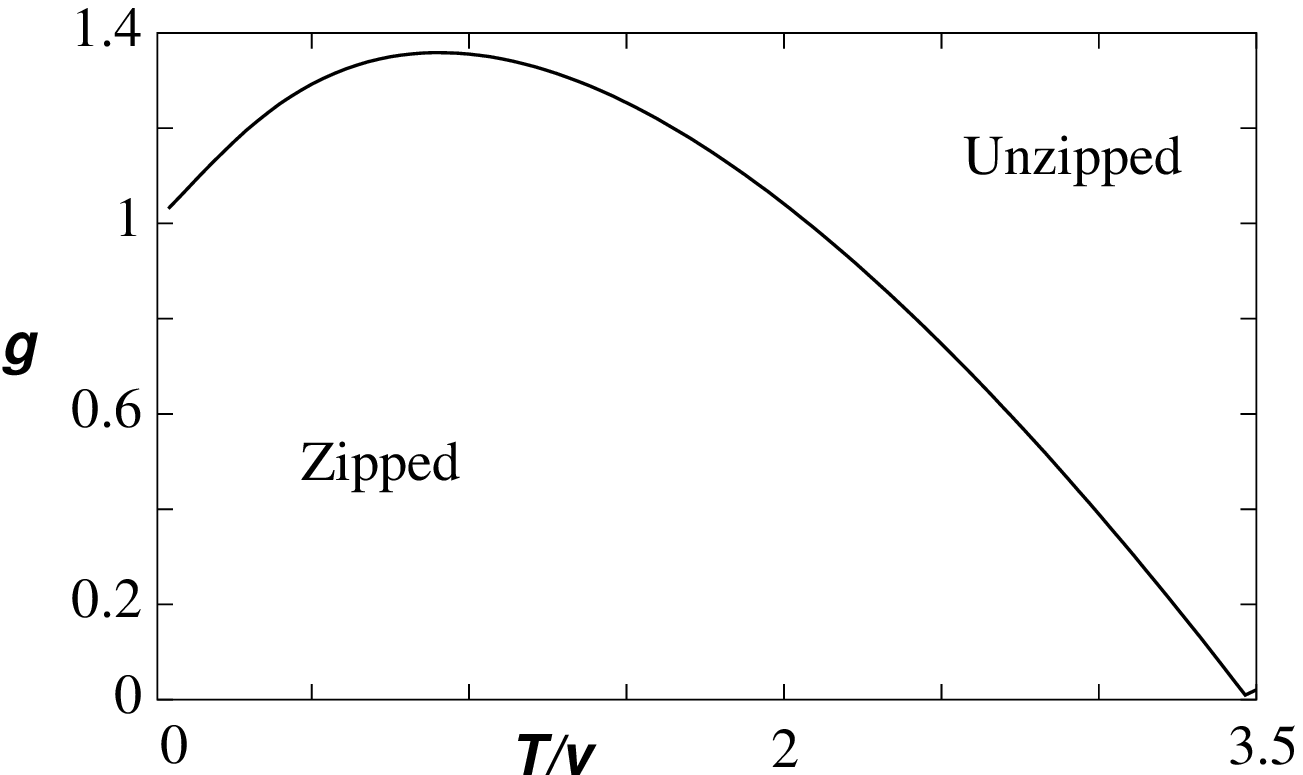} 
   \caption{Unzipping
   phase diagram (exact) for two mutually avoiding directed polymers
   with an attractive interaction $v$ in $1+1$ dimensions. Note the
   reentrance at low temperatures. From Ref. \cite{maren1}.  } 
  \label{fig:dna}
 \end{figure}
}

\newcommand{\figsteps}{%
\begin{figure}[htbp]
   \centering
   \includegraphics[scale=0.6]{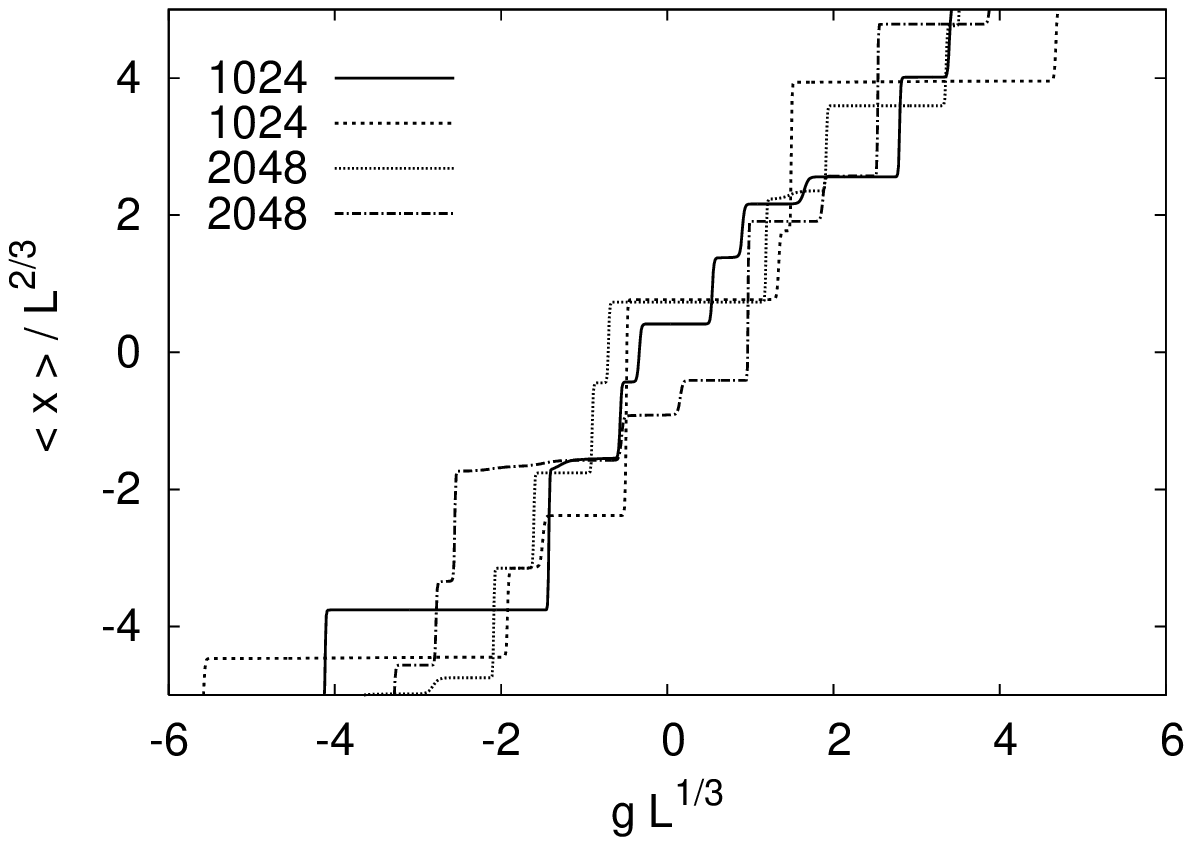}
   \includegraphics[scale=0.6]{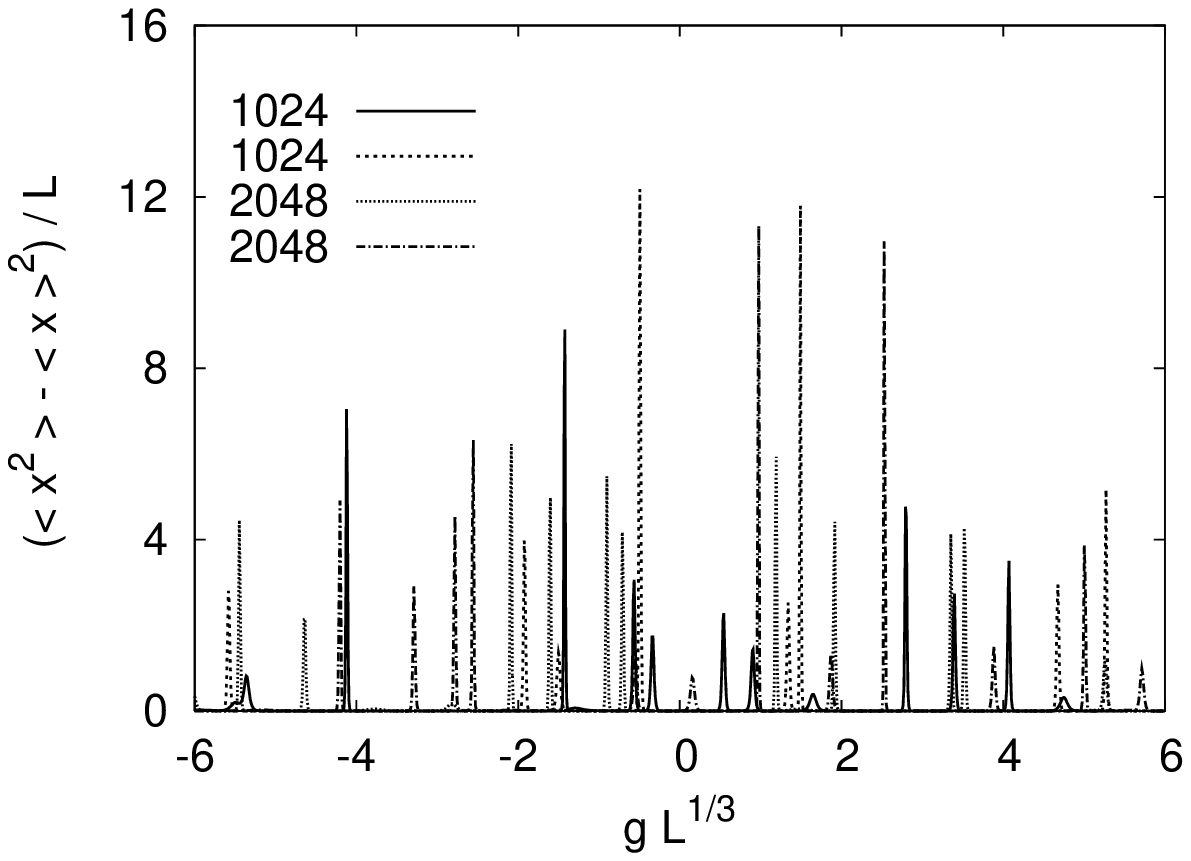}

   \caption{ (a) Plot of $<r>/N^{\nu}$ vs $gN^{\nu-\theta}$ for $d=1$.
   Two different values of $N$ and in different realizations of
   disorder. (b) Corresponding fluctuation in position.  From
   Ref. \cite{rajeev}.} \label{fig:step} 
\end{figure} 
}

\newcommand{\figrani}{%
\begin{figure}[htbp]
   \centering
   \includegraphics[scale=0.6]{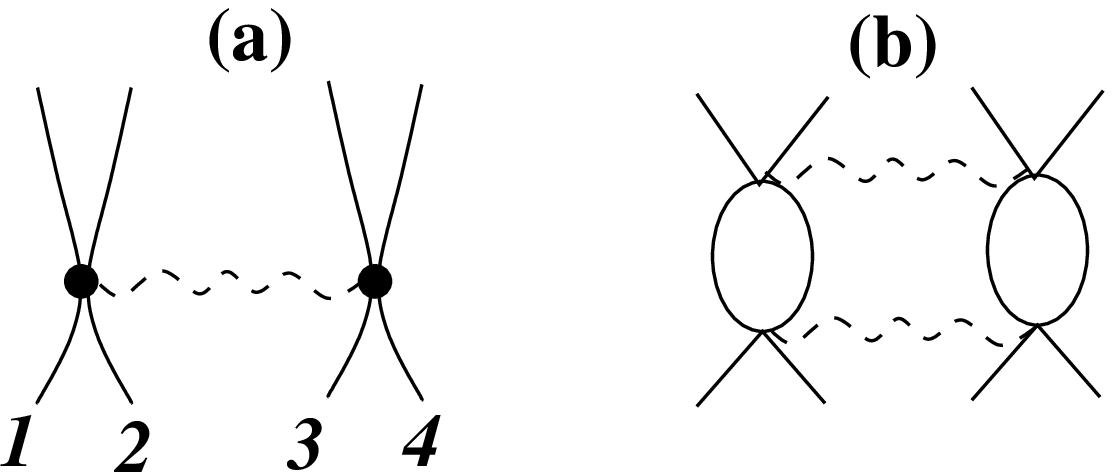}

   \caption{ (a) Inter-replica interaction in the RANI model.  The
     dotted wavy line indicates the ``$r$''-type interaction between
     the pairs (1,2) and (3,4).  (b) indicates a loop formed by the
     disorder induced interaction.
      }
   \label{fig:rani}
 \end{figure}
}

\newcommand{\figolp}{%
\begin{figure}[htbp]
   \centering
   \includegraphics[scale=0.6]{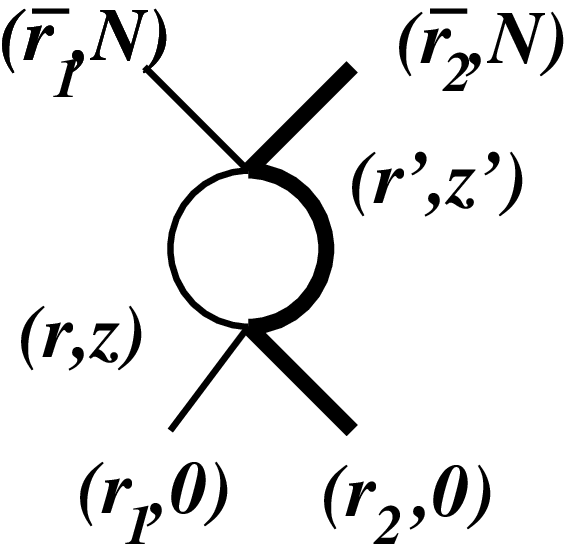}
   \caption{  The one loop diagram for two polymers.  The labels
     denote the position and the $z$ of the points.  There are
     integrations over all these free coordinates.
     }
   \label{fig:olp}
 \end{figure}
}

\newcommand{\figapp}{%
\begin{figure}[htbp]
   \centering
   \includegraphics[scale=0.6]{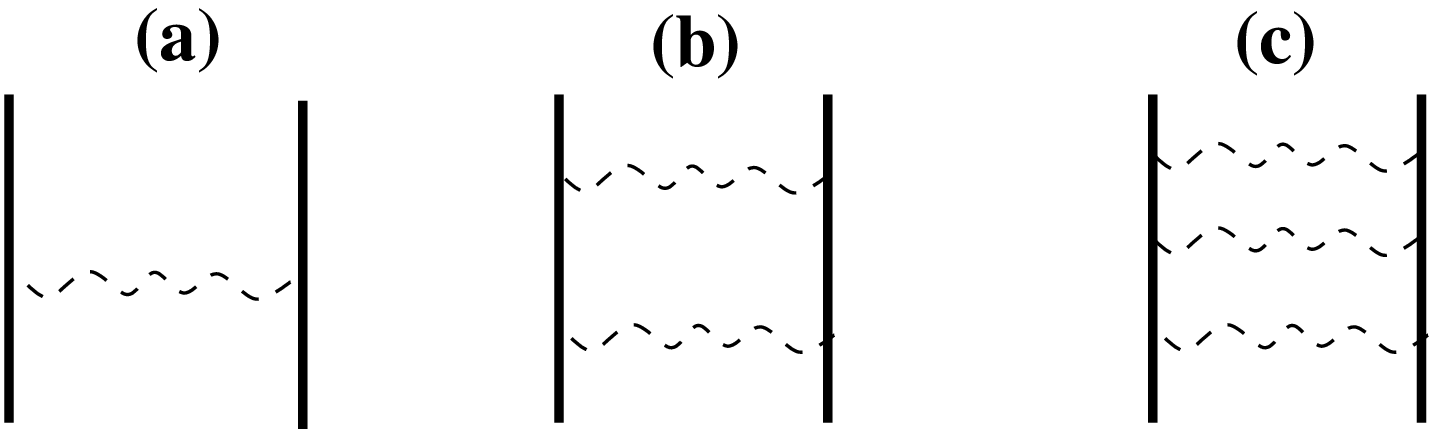}
   \caption{ (a) The contributing ladder diagrams for $\dav[c]{Z^2}$.
     A thick line corresponds to a pair of chains. A wiggly line
     stands for an ${\bar{r}}_0$ factor in the evaluation of the
     diagrams.  There are one and two loops in diagrams (b) and (c)
     respectively.  Divergences arise from loop integrations.  }
   \label{fig:app}
 \end{figure}
}

\newcommand{\figtrf}{%
\begin{figure}[htbp]
   \centering
   \includegraphics[scale=0.6]{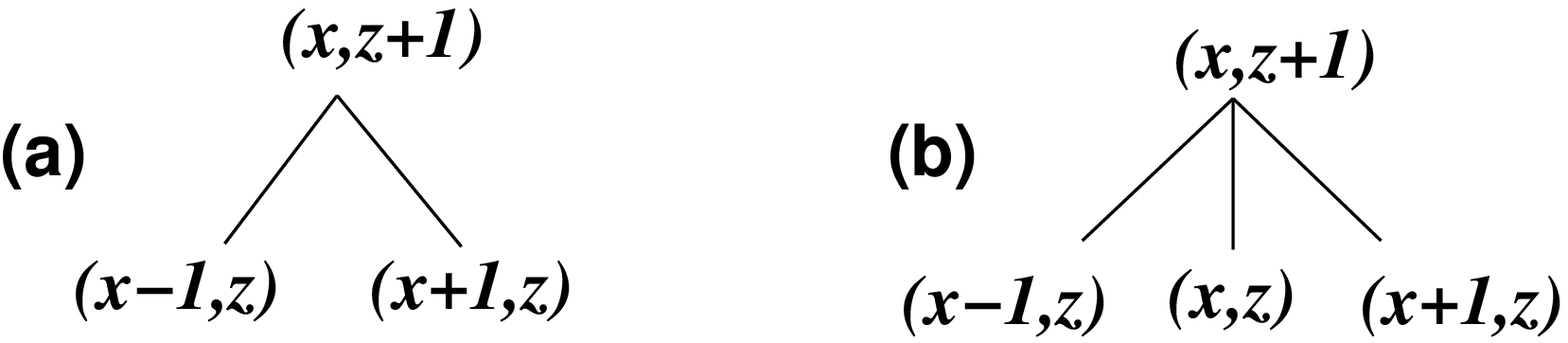}
   \caption{ Lines show the possible steps from step $z$ to $z+1$
     on a squarre lattice.  Backward steps are not allowed for
     directed polymers.  (1) Lattice oriented by 45 degrees.  The $x$
     and $z$ axes are along the diagonals of the square lattice.  (b)
     Axes are along the two directions of the square lattice but the
     polymer can take steps along the diagonals of the unit cell also.
   }
   \label{fig:trf}
 \end{figure}
}

\newcommand{\figflw}{%
\begin{figure}[htbp]
   \centering
   \includegraphics[scale=1]{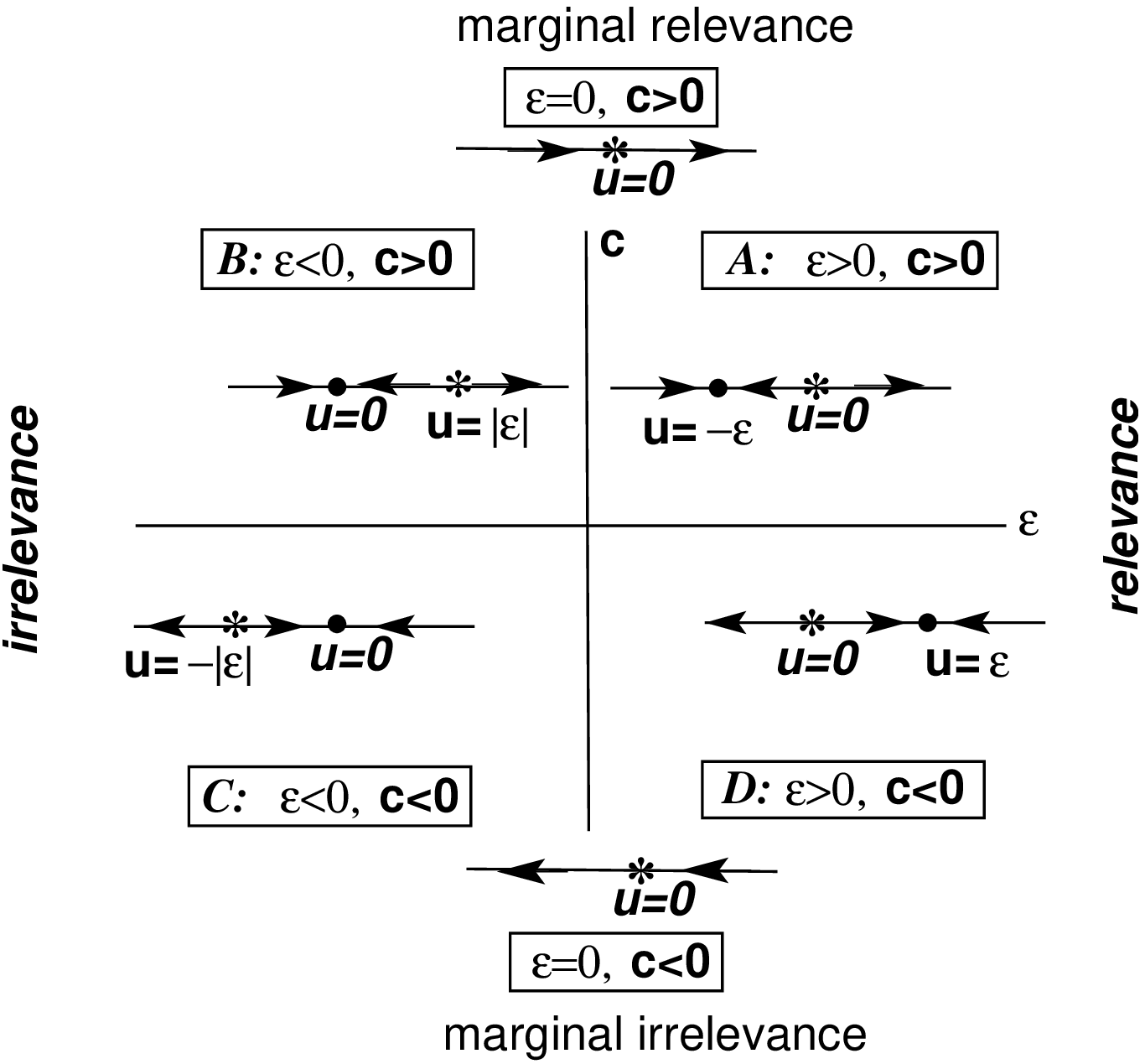}
   \caption{ The various types of flow diagrams depending on
     $\epsilon$ and $c$, the coefficients of the linear and the
     quadratic terms of the flow equation.  Solid bullet ($\bullet$) represents a
     stable fixed point while the star ($*$) represents an unstable
     fixed point.
   }
   \label{fig:flw}
 \end{figure}
}

\begin{document}

\title{Directed polymers and Randomness}
\author{Somendra M. Bhattacharjee\\
Institute of Physics, Bhubaneswar 751 005, India\\ email:
somen@iopb.res.in }
\maketitle

\begin{abstract}
The effects of two types of randomness on the behaviour of directed
polymers are discussed in this chapter. The first part deals with the
effect of randomness in medium so that a directed polymer feels a
random external potential.  The second part deals with the RANI model
of two directed polymers with heterogeneity along the chain such that
the interaction is random. The random medium problem is better
understood compared to the RANI model.
\end{abstract}


\section{Directed polymers}

A long flexible elastic string, to be called a polymer, has several
features of a critical system, showing power law behaviours without
much fine tuning \cite{degen,doi}.  An important quantity for a polymer
is its size or the spatial extent as the length $N$ becomes large.
For a translationally invariant system with one end ($z=0$) fixed at
origin, the average position at $z=N$ is zero but the size is given by
the rms value
\begin{equation}
  \label{eq:9}
  \langle {\bf  r}_N\rangle=0,\quad \langle r^2_N\rangle ^{1/2}\sim 
  N^{\nu},
\end{equation}
with $\nu=1/2$, for the free case.  Similar power laws can be defined
in other properties also.  In general, such exponents
like the size exponent $\nu$ define the polymer universality
class
and these depend only on a few basic
elements of the polymer.  In addition to the geometric properties, the
usual thermodynamic quantities, e.g.  free energy (or energy at
temperature $T=0$), entropy etc., are also important, especially if one wants to
study phase transitions.

Power laws
generally imply absence of any length
scale in the problem.  Given a microscopic Hamiltonian with its own
small length scales like the bond length, interaction range etc, power
laws occur only in the long distance limit (large $N$) for thermally
averaged 
quantities which require summing
over all possible configurations.  As a result, in the long distance
limit, these power laws become insensitive to minute details at the
microscopic level, a feature that may be exploited to choose
appropriate simplified models to describe a polymer.

In thermal equilibrium, the Boltzmann
distribution
ultimately determines the
macroscopic behaviour.  In most cases one may avoid the issue of
probability distribution
but instead
may concentrate only on the first few, may be the first two, moments
or cumulants.  For example, one needs to know the average energy,
entropy etc and the various response functions which come from the
width of the distribution.
Thermodynamic
descriptions do not generally go beyond that.

In random physical systems, one encounters an extra problem. If the
randomness is not thermalizable
(``quenched''),
any quantity of interest
becomes realization dependent. As a result, an additional disorder
averaging
has to be done over and above the
usual thermal averaging for each realization.  It is therefore
necessary to know if and how this extra averaging alters the behaviour
of the system.

Critical-like systems \cite{crit} become a natural choice for studying
the effect of quenched randomness because it is hoped that at least
some of the properties would be insensitive to the minute details of
the randomness.  Since for a critical system, the influence of the
randomness on a long scale is what matters, it transpires that the
critical behaviour will be affected if the disorder is a relevant
variable.  In the renormalization group language,
a coupling is classified as {\it relevant, irrelevant or marginal},  if,
with increasing length scale, it grows, decays or remains invariant,
because the contribution of a relevant quantity cannot be ignored at
long distances even if it is numerically small to start with. 

With relevant disorder, there is the obvious possibility of a change
in the critical properties (e.g. new set of critical exponents). More
complex situations may include emergence of new or extra length
scales.  One may recall that approach to criticality, most often, is
described by a diverging length scale.  Developing a description of
the system in terms of this large length scale only goes by the name
of {\it scaling} theory.  Emergence of any new or extra length scale
would then alter the corresponding scaling description. In case it is
possible to change the nature of disorder from relevant to irrelevant
(say by changing temperature), then a phase transition would occur
which will have no counterpart in the pure problem.  For the
disorder-dominated phase, on a large scale, there are possibilities of
rare events (see Appendix \ref{app:typ}) which necessitates a
distinction between the average value and the typical (e.g.  most
probable) value.  In such situations higher moments of the quantity
concerned become important.  These are some of the aspects that make
disorder problems important, interesting and difficult.

\figone 

The problem of a polymer in a random medium was initiated by
Chakrabarti and Kertesz \cite{bkc,krish} with the application of the Harris
criterion.
This problem has enriched our
overall understanding of polymers and random systems in general, but
still a complete understanding remains elusive.  Not surprisingly,
the look out for  simpler problems that capture the basic essence of the
original complex system gained momentum.  In this context, directed
polymers
played a very crucial role.
 
Let us define the problem here. Consider a polymer where each monomer
sees a different, independent, identically distributed random
potentials.  Geometrically this can be achieved if the monomers live
in separate spaces.  One way to get that is to consider the polymer to
be a $d+1$ dimensional string with the monomers in $d$ dimensional
planes but connected together in the extra dimension.  As shown in
Fig. \ref{fig:1}, this is a polymer which is directed in one
particular direction.  Hence the name directed
polymer \cite{hhz,kard2,hhf,HF}.

For a directed polymer, the size would now refer to the size in the
transverse $d$-directions and so Eq. (\ref{eq:9}) refers to the
transverse size as the length in the special $z$-direction increases
For long enough chains, it is this size that matters and enters the
scaling description.

The significance of directed polymer lies in the fact that the pure
system is very well understood and exactly solvable in all dimensions
while the random problem can be attacked in several different ways, a
luxury not affordable in most situations.

Two types of randomness can arise in the context of directed polymers.
One type would involve imposition of a random external potential
(random medium problem).  In the second type, the interaction (between
say two chains) is random (RANI model).  In the random medium problem,
the random potential would like to have a realization dependent ground
state which may not match with the zero-field state.  In the RANI
model, the randomness in the interaction may lead to a change in the
phase transition behaviour shown by the polymers. These two classes
are discussed separately.

\section{Outline}
We first consider the random medium problem and then the random
interaction (RANI) model.  In both cases, disorder turns out to be
marginally relevant though at two different dimensions.  The
quantities of interest in a disordered system are introduced in Sec.
\ref{sec:hamilt-rand}.  For the random medium problem, the question of
relevance of disorder, the size exponent $\nu$ and the free energy
fluctuation exponent $\theta$ are discussed in Secs.
\ref{sec:disorder-relevant} and \ref{sec:can-we-handle}.  Sec
\ref{sec:disorder-relevant} deals with the renormalization group (RG)
for the moments of the partition function and also the special Bethe
ansatz results for the $1+1$ dimensional problem.  A Flory
approach and scaling ideas are also summarized there.  Sec.
\ref{sec:can-we-handle} deals with the RG approach for the free energy
via the Kardar-Parisi-Zhang equation.  The behaviour of the overlap
especially near the transition to strong disorder phase may be found
in Sec. \ref{sec:over} .  We briefly mention the unzipping behaviour
in presence of a force and summarize some recent results for the pure
case in Sec. \ref{sec:unz_pure}.  More on unzipping of a two chain
system may be found in Sec. \ref{sec:unzip}.  These results and scaling
arguments are then used to visualize the nature of the ground state in
Sec. \ref{sec:intuitive-picture}.  Various aspects of the RANI model can 
be found in Secs.  \ref{sec:rani} and \ref{sec:hier}.  The question of 
relevance, and 
annealed versus quenched disorder in multi chain system are analyzed
in Sec. \ref{sec:rani}.  The two different types of randomness or
heterogeneity on hierarchical lattices are studied in Sec.
\ref{sec:hier}.  In the last part of this section, one may find the
general validity and extension of the Harris criterion
as applied to this polymer problem.  Various technical
issues are delegated to the Appendixes.  An example of rare events is
given in App. \ref{app:typ}.  A short review of the pure polymer
behaviour can be found in App. \ref{app:pure}.  The issue of self-averaging
and some recent results about it for disordered systems are
surveyed in App.  \ref{app:sa}.  The renormalization group approach to
polymers as used in Sec.  \ref{sec:zrg} is spelt out in App.
\ref{app:rg} in the minimal subtraction scheme with dimensional
regularization. The momentum shell RG approach for the free energy is
dealt with in App.  \ref{app:kpz}.  All the possible flow diagrams are
reviewed in App.  \ref{app:flwec}.  A short introduction to the
transfer matrix approach for the directed polymer problem is given in
App.  \ref{app:transf}.  The exact RG for the RANI model can be found
in App. \ref{app:rani}.

{\bf On Notation and convention:} 
\begin{itemize}
\item To avoid proliferation of symbols, we reserve the
symbol ${\sf f}$ to denote an arbitrary  or unspecified function, not
necessarily same everywhere.
\item  By a {\it sample} or a {\it realization}  we would mean one particular 
arrangement of the random parameters over the whole system.  For a
thermodynamic (infinitely large) system the sample space is also infinite.
\item  {\it Sample averaging} is denoted by $\dav{...}$
while {\it thermal averaging} is denoted by
$\langle...\rangle$.
\item The Boltzmann constant is set, most often, to one, $k_{\rm
    B}=1$.
\item ``Disorder'' and ``randomness'' will be used interchangeably.

\end{itemize}

\section{Hamiltonian and Randomness}
\label{sec:hamilt-rand}

By definition, a directed polymer is defined in $D$ dimensions out of
which one direction is special that represents the sense of direction
of the polymer.  It is then useful to show that explicitly by writing
$D=d+1$ where $d$ is the dimension of the transverse space. One may
consider various possible models of the underlying space as shown in
Fig. \ref{fig:1}.
\begin{itemize}
\item   One may consider a lattice (square lattice in the
Fig. \ref{fig:1}b) with the polymer as a random walk on the lattice
with a bias in the diagonal $z$-direction, never taking a step in the
$-z$ direction. The length of the polymer is then the number of steps
on the lattice. 
\item Instead of a lattice model, one may consider continuum
where both the space and the $z$-direction are continuous as shown in
Fig.  \ref{fig:1}c.  The polymer itself may consist of small rods or
bonds whose length provides us with a small scale cut-off.  In many
situations, it helps to consider polymers without any intrinsic small
scale cutoff. 
\item Quite often it is useful to consider very special
lattices, e.g. hierarchical lattices 
as shown in Figs. \ref{fig:1}d, and \ref{fig:1}e, because of the possibility 
of exact analysis.  Here one starts with a bond and then replaces the bond
iteratively by a predetermined motif (``diamond'' of $2b$ bonds) and
the process can be iterated {\it ad infinitum}. This is a lattice with
a well-defined dimension (see Sec. \ref{sec:hier}) but unfortunately
without any proper Euclidean distance. Consequently geometric
properties are not useful here.  The effective dimension of the
lattice is $d_{\rm eff}=(\ln 2b)/\ln 2$, if there are $2b$ bonds per
motif.  A directed polymer can be taken as a random walk on this
lattice starting from say the bottom point, going up, and ending at
the top end.
\end{itemize}

\subsection{Pure case}
\label{sec:pure-case}
Taking the polymer as an elastic string, one may define a Hamiltonian
\begin{equation}
\label{eq:1.1}
H_0=\frac{d}{2} K \int dz \left( \frac{\partial {\bf r}}{\partial
z}\right ) ^{^2}
\end{equation}
which gives a normalized probability distribution
of the position vector ${\bf r}$ at length $z$ from the end at $({\bf
  0},0)$ 
\begin{equation}
\label{eq:1.2}
P({\bf r},z) = \frac{1}{(2\pi z)^{d/2}} \ e^{- r^2/2z} \qquad
(Kd/k_{\rm B}T =1).
\end{equation}
Here $k_{\rm B}$ is the Boltzmann constant and $T$ is the temperature.
One can even write down the distribution for any two intermediate
points $({\bf r}_i,z_i)$ and $({\bf r}_f,z_f)$ as 
\begin{equation}
  \label{eq:60}
   G({\bf r}_f,z_f|{\bf r}_i, z_i) = \frac{1}{[2\pi (z_f-z_i)]^{d/2}}
 \  \exp\left(- \frac{({\bf r}_f-{\bf r}_i)^2}{2(z_f-z_i)}\right)
\end{equation}
For the lattice random walk, there is no ``energy'' and the elastic
Hamiltonian of Eq. (\ref{eq:1.1}) just simulates the entropic effect at
non-zero temperatures. One needs to look at the lattice problem in
case one is interested in low or zero temperature behavior.  A
recapitulation of a few properties of polymers is done in Appendix
\ref{app:pure}.

For a polymer of length $N$ the probability distribution gives
\begin{equation}
\label{eq:1.3}
\langle {\bf r}\rangle=0, \quad \langle  r^2\rangle= N, \qquad (K d/k_{\rm B}T=1)
\end{equation}
so that the transverse size of the polymer is given by 
\begin{equation}
\label{eq:1.3a}
R_0 \sim\langle  r^2\rangle^{1/2}\sim N^{\nu}\ {\rm with}\ \nu=1/2.
\end{equation}
The power law growth of the size of a polymer as the length increases
is a reflection of the absence of any ``length scale'' in the
Hamiltonian.

\subsection{Random medium}
\label{sec:hamiltonian}
Let us now put this polymer in a random medium.  In the lattice model
of Fig. \ref{fig:1}, each site has an independent random energy and
the total energy of the lattice polymer is the sum of the energies of
the sites visited.  In continuum, the Hamiltonian can be written as
\begin{equation}
\label{eq:2.1.1a}
H=H_0
+ \int_0^N dz\   \int d{\bf r}\  \eta({\bf r},z)\ \delta({\bf r}(z)-{\bf
  r}) 
=   H_0+ \int_0^N dz\ \eta({\bf r}(z),z)
\end{equation}
where $\eta({\bf r},z)$ is an identical, independent Gaussian
distributed
random variable with zero
mean and variance $\Delta>0$,
\begin{equation}
  \label{eq:3}
\dav{\eta({\bf r},z)}=0, \  \dav{\eta({\bf r},z)
 \ \eta({\bf r}^{\prime},z^{\prime})}
=\Delta\delta({\bf r}-{\bf r}^{\prime})\delta(z-z^{\prime}) .
\end{equation}
The averaging over $\eta$ is to be called {\it sample averaging},
denoted by $\dav{...}$ (as opposed to thermal
averaging, denoted by $\langle...\rangle$).  With this distribution of
random energies, we see $\dav{H}=H_0$ and so the average Hamiltonian
is not of much use.  Disorder averaging of sample dependent thermal
averages is to be called {\it quenched averaging} while disorder
averaging done at the partition function level is to be called {\it
annealed averaging}. 

We shall consider the situation with one end point ($z=0$) fixed.
Otherwise, the polymer may drift in the medium to locate the best
possible region that would minimize its free energy.  Such a case,
eventhough formally quenched in nature, is tantamount to an annealed
case.

\subsubsection{Partition function}
\label{sec:partition-function} 
The partition function for a polymer in a random medium or potential
is given by
\begin{equation}
\label{eq:2.1.1}
Z=\int {\cal DR} \ e^{-\beta H}.
\end{equation}
This is a symbolic notation (``path integral'')
to denote sum over all configurations and
is better treated as a continuum limit of a well-defined lattice
partition function
\begin{equation}
\label{eq:2.1.2}
Z=\sum_{\rm paths} e^{-\beta \eta({\bf r},z)}
\end{equation}
where the sum is over all possible paths of $N$ steps starting from
${\bf r}={\bf 0}$ at $z=0$.  It often helps to define the partition
function such that $Z(\{\eta=0\}) = 1$ to avoid problems of going to
the continuum limit (see Eq. (\ref{eq:1.2})).  This is done by dividing
(or normalizing) $Z$ by $Z_0=\mu^N$, $Z_0$ being the partition
function of the free walker with $\mu$ as the connectivity
constant
($=2$ for Fig. \ref{fig:1}b).

Let us define the free energy 
\begin{equation}
  \label{eq:24}
F=-T\ln Z,  
\end{equation}
for a polymer of length $N$ where the end point at $z=N$ is free.  A
more restricted free energy may be defined as
\begin{equation}
  \label{eq:25}
 F({\bf r},N) =-T\ln Z({\bf r},N), 
\end{equation}
when the end at $z=N$ is at ${\bf r}$.
 
\subsection{Unzipping and response}
Quite often it is useful to force a system to change its overall
configuration by applying an external field
The response function then tells us about the rigidity of the
system against such external perturbations.  E.g.,  a magnetic field
may be applied on  a magnet and the magnetic susceptibility is the
corresponding response function.  
A similar applied force for directed polymer is an unzipping
force or a pulling force applied at one end (see Fig. \ref{fig:frc}).
We call such a force an unzipping force
because of its role in unzipping of DNA-type double stranded
polymers \cite{smb_mt}.  There could two different ensembles.  One is a
fixed force ensemble where one applies a force at the free end $z=N$
and studies the change in the size and shape of the polymer or its
response.  The position of the end point is given by $r = - T \partial
F(g,N)/\partial g$.  The second is the fixed distance ensemble where
the free end is at point ${\bf r}$ and then what is the force required
to maintain it at that point.  Using the constrained free energy, we
may write ${\bf g}=-T
\nabla F({\bf r},N)$.  (Here we used the same notation $F$ to denote
the free energies of the two ensembles. The arguments and the context
would distinguish the two.)  The two ensembles behave differently in a
disordered system.\footnote{ The inequivalence of the two ensembles is
known also for pure case if the force is applied at some intermediate
point \cite{kap04}}.

\figfrc

If we consider the response
of a directed polymer to the unzipping force, the response function
comes from the Hamiltonian
\begin{eqnarray}
\label{eq:2.3.1}
H&=& \frac{d}{2}K \int dz \left( \frac{\partial
{\bf r}}{\partial z}\right ) ^{^2} + \int dz\  \eta({\bf r}(z),z) - {\bf
g}\cdot \int \frac{\partial 
{\bf r}}{\partial z} \ dz \nonumber\\
&=&\frac{d}{2} K\int dz \left( \frac{\partial}{\partial z} ({\bf r} -
\frac{{\bf g}z}{Kd}) \right ) ^{^{\scriptstyle 2}} + \int dz \ \eta({\bf r}(z),z) 
-\quad\frac{1}{2} \frac{g^2 N}{Kd}.
\end{eqnarray}
The disorder is Gaussian-distributed as in Eq. (\ref{eq:3}).  
The general response function for the force is
\begin{equation}
\label{eq:2.3.2}
 \left . C_T\right|_{_{ij}} = \left.\frac{\partial^2 \dav{\ln
Z}}{\partial g_i\partial g_j}\right |_{_{g=0}} = \dav{\langle r_i
r_j\rangle - \langle r_i\rangle \langle r_j\rangle},
\end{equation}
with $i,j$ representing the components.  It is known in statistical
mechanics that the response of a system in equilibrium is determined
by the fluctuations.

\subsubsection{Exact result on response: pure like}
\label{sec:exact-results}
By a redefinition of the variables and using the $\delta$-correlation
of the disorder in the $z$ direction, we
have
\begin{equation}
\label{eq:2.3.3}
\dav{\ln Z({\bf g})} = \dav{\ln Z({\bf g}=0)} + \frac{g^2 N}{Kd}, 
\end{equation}
from which it follows that
\begin{equation}
\label{eq:2.3.4}
 C_T = \ \frac{TN}{Kd},
\end{equation}
as one would expect in a pure system, Eq. (\ref{eq:1.2}).  And there
are no higher order correlations.

Two things played important roles in getting this surprising pure-like
result: (i) The disorder correlation has a statistical translational
invariance coming from the delta function in the $z$-coordinate, and
(ii) the quadratic nature of the Hamiltonian.  If disorder had any
correlation along the length of the polymer, Eq. (\ref{eq:2.3.4})  will
not be valid.

The significance of the result is that the conventional thermal
fluctuation, averaged over randomness, superficially does not say much
about the effect of disorder.  We shall see later that this innocuous
result however contains important information.

\subsubsection{Quantities of interest}
\label{sec:averages}
Let us list some of the quantities which are of interest for a
disordered system. 

\begin{itemize}
\item A random system needs to be described by the probability distribution
of various physical quantities or by the averages and moments (over
realizations). The moments are useful, especially in absence of full
information on the probability distribution and also for
characterization of the properties of the polymer.  Since there is no
unique partition function, one of the important probability
distributions would be of the partition function, $P(Z)$.  Any
quantity of interest needs to be averaged over such a distribution.
Similarly the probability distribution $P(F)$ of the free energy is
also of interest.  The thermodynamic behaviour is obtained from
$\dav{F}$. In case the probability distribution (over the realizations)
of a thermal averaged quantity $X$ becomes very sharp, especially in
the large size limit, one may avoid doing the disorder averaging.
This may happen for extensive quantities because of additivity over
subsamples.  Such quantities are called self-averaging. Certain
aspects of self-averaging is discussed in Appendix \ref{app:sa}.

\item The first thing to determine  is the relevance of disorder.
To do so,  we may write 
\begin{equation}
  \label{eq:2}
\dav{\ln Z} = \dav{\ln \{\dav{Z} + (Z-\dav{Z} )\} } =
\ln \dav{Z} +\frac{\dav{Z^2} -\dav[2]{Z} }{2\dav[2]{Z}}  +... .
\end{equation}
This shows the importance of the variance of the partition function.
If the variance remains small, in the limit $N\rightarrow\infty$, then
the polymer can be described by the average partition function which
is more or less like a pure problem. Otherwise not.  We see that the
relevance of the disorder may be inferred from the nature of the
variance of the partition function.

\item For the partition function we may use the simple identity
\begin{equation}
\label{eq:rep.1}
\dav{Z^n} = \dav{e^{n\ln Z}} = \exp \left( \sum 
\frac{1}{m!}n^m \dav[(c)]{(\ln Z)^m}\right ),
\end{equation}
where $\dav[(c)]{(\ln Z)^m}$ are the cumulants.  In contrast to
Eq. (\ref{eq:2}), it is now the fluctuations of the free
energy
that become important.  One may introduce a scaling behaviour, namely
\begin{equation}
  \label{eq:7} 
\dav[(c)]{(\ln Z)^2} \equiv \dav{(\ln Z)^2} -\dav{\ln  Z}^2 \sim N^{2\theta},
\end{equation}
defining a new exponent $\theta$.
Obviously,
for a pure problem $\theta=0$.  If higher order fluctuations (or
cumulants) do not require any new exponent, then it is fair to expect
$\dav[(c)]{(\ln Z)^m}\sim N^{m\theta}$. This free energy fluctuation
exponent is one of the new quantities required to describe the random
system.

\item A simpler form of Eq. (\ref{eq:rep.1}) is the basis of the
replica
approach for disordered systems, namely
\begin{equation}
\label{eq:repl}
  \dav{\ln Z} =\lim_{n\rightarrow 0} \frac{\dav{Z^n} - 1}{n}.
\end{equation}
so that to compute the average free energy we may consider a case of
$n$-replicas of the original system or after averaging, an $n$-polymer
problem with extra interactions induced by the disorder though an
$n\rightarrow 0$ limit is to be taken at the end.  A few possible
paths to take the limit for long chains are shown in
Fig. \ref{fig:nN}.

\fignN

Nontrivial results are expected if and only if the origin in Fig.
\ref{fig:nN} is a singular point so that the limits
$n\rightarrow 0$ and $N\rightarrow\infty$ become non-interchangeable.
In other words, the $n$ and $N$ dependences should be coupled so that
the appropriate path is a scaling path like (b) in the figure
\ref{fig:nN}.

\item If we demand that $\ln \dav{Z^n}$ is proportional to $N$ for large
$N$, then, apart from the extensive term ($\propto nN$), there will be
corrections which may be assumed to involve a scaling variable
$x=nN^{\theta}$.  For $x\rightarrow\infty$, ${\cal F}(x) \sim
x^{1/\theta}$ so that
\begin{equation}
\label{eq:rep.3}
\ln \dav{Z^n} = n \epsilon N + {\sf a} n^{1/\theta} N. \ (N\rightarrow\infty)
\end{equation}
This is for path (a) of Fig. \ref{fig:nN}.  In contrast, if we take
$n\rightarrow 0$ for finite $N$, path (c), a Taylor series expansion
gives
\begin{equation}
\label{eq:pathc}
\ln \dav{Z^n} = n N\epsilon + {\sf a} n^2 N^{2\theta} + ... .\quad
(n\rightarrow 0)
\end{equation}
Eqs. \ref{eq:rep.3},\ref{eq:pathc} can be used to calculate $\theta$,
the free energy fluctuation exponent.

\item So far as the geometrical properties are concerned, we first note the
lack of translational invariance for a particular realization of
disorder and therefore $\langle {\bf r}\rangle \neq 0$, but on
averaging over randomness, translational invariance will be restored
statistically and so $\dav{\langle {\bf r}\rangle}$ has to be zero.
One may therefore consider the size of the polymer by the ``disorder''
correlations
\begin{equation}
\label{eq:10}
C_{\rm dis}\equiv \dav{\langle {\bf r}^2\rangle}\ {\rm or}
\  C_{\rm dis}\equiv \dav[2]{|\langle   {\bf r}\rangle|}\ {\rm with} 
\ C_{\rm dis}\sim N^{2\nu}
\end{equation}
We have already seen in Eq. (\ref{eq:2.3.4}) that the usual
correlation function is the thermal correlation of the pure problem
and has no signature of the disorder.  Of course, the disordered
averaged probability distribution $P( {\bf r},N)=\dav{Z({\bf
r},N)/\int d{\bf r} Z({\bf r},N)}$ is also of
importance.

\item For the pure case ($\Delta=0$), there is no ``energy'', only
configurational entropy of the polymer.  But with $\Delta \neq 0$,
there may be one or more lowest energy states.  The nature of the
ground state is an important issue for any disordered problem.  For
the lattice problem, the energy is the sum of $N (\rightarrow\infty)$
random energies of the sites visited.  But it is the minimization over
a connected path that makes the problem difficult.  If the low
temperature behaviour of the polymer is same as that at $T=0$, the
fluctuation of the ground-state energy will also be described by the
exponent $\theta$ of Eq. (\ref{eq:7}).  Same is true for the size
exponent also.  Such a situation requires that the disorder dominated
phase is controlled by the ``zero temperature'' fixed point.

\item The problem we face in a disordered system is that there is no well
defined ground state - the ground state is sample dependent.  There is
therefore no predefined external field that will force the system into
its ground state (e.g., a magnetic field in a ferromagnetic problem).
This is a generic problem for any random system.

But, suppose, we put in an extra fictitious (ghost) polymer and let it
choose the best path.  Now we put in the actual polymer in the same
random medium but with a weak attraction $v$ with the ghost. At $T=0$,
this polymer will then sit on top of the ghost.  In absence of any
interaction ($v\rightarrow 0$ ), the polymer would go over the ghost
in any case if there is a unique ground state, not otherwise.  At
non-zero temperatures there will be high energy excursions and how
close to the ground state we are will be determined by the number of
common points of the two polymers.  This is called overlap which may
be quantitatively defined as
\begin{equation}
\label{eq:2.4.1}
q_i=\frac{1}{N} \int \delta({\bf r}_1(z)-{\bf r}_2(z)),
\end{equation}
for a given sample $i$ and then one has to average over the disorder
samples, $q\equiv\dav{q_i}$.

In case of a repulsive interaction, the situation will be different.
If there is more than one ground state, the two chains will occupy
two different paths and there is no energetic incentive to collapse on
top of each other when the repulsion $v\rightarrow 0+$.  In such a
scenario, the overlap $q(v\rightarrow 0+)\neq q(v\rightarrow 0-)$.
Conversely, a situation like this for the overlap would indicate
presence of degenerate ground states.  For a unique ground state, the
second chain would follow a nearby excited path with certain amount of
overlaps with the ground state.  A little reflection shows that
overlap is associated with the second moment of the partition
function.
\end{itemize}

\section{ Relevance of disorder}
\label{sec:disorder-relevant}
Let us first see if disorder is at all relevant.
 
\subsection{Annealed average: low temperature problem}
If the disorder is irrelevant, then the average partition function
$\dav{Z}$ (annealed average) is expected to give the thermodynamic
behaviour.  However this extra averaging of the partition function may
lead to a violation of the laws of thermodynamics, questioning the
validity of the annealed averaging.  This happens for directed
polymers.  With a Gaussian distribution for the random energy, from
Eq. (\ref{eq:2.1.2}),
\begin{eqnarray}
\label{eq:2.1.5}
\dav{Z}=  \exp(\beta^2\Delta N/2) \exp(N\ln \mu),
\end{eqnarray}
so that $F/N = -T( \ln \mu + \beta^2 \Delta/2)$.  The entropy obtained
from this partition function ($S=-\partial F/\partial T$) by
definition has to be positive which requires $T\geq T_{\rm A}\equiv
\sqrt{\frac{\Delta}{2\ln\mu}}$.  Annealed averaging will not work at
very low temperatures.

This does not necessarily mean that something like a phase transition
has to happen, because this problem will occur for any disordered
system, even finite in size.   However, for the  directed
polymer problem, this does signal a phase transition, though the proof
comes from other analysis.

\subsection{Moments of $Z$}
The moments can be written as the partition function of an $n$-polymer
problem with an extra interaction induced by the disorder.  Starting
from $H$ as given by Eq. (\ref{eq:2.1.1a}), and averaging over the
Gaussian distribution of Eq. (\ref{eq:3}), we have
\begin{eqnarray}
  \label{eq:4} 
  \dav{Z^n} &=&\int {\cal DR}_1\ {\cal DR}_2...{\cal DR}_n 
   \ e^{-\beta H_n},\nonumber\\ 
{\rm where\ } H_n&=& \frac{Kd}{2}
  \int_0^N dz \sum_i^n\left( \frac{\partial {\bf r}_i}{\partial z}\right ) ^{^2} 
   - \beta \Delta \int_0^N d z \ \sum_{i<j} \delta({\bf r}_i(z)-{\bf r}_j(z)).
  \label{eq:5}
\end{eqnarray}
This particular form can be understood in terms of two polymers, to
which one may restrict if the interest lies only in the second
cumulant of $Z$.  These two polymers start from the same point and do
their random walk as they take further steps.  If there is a site
which is energetically favourable, both the polymers would like to be
there.  The effect is like an attractive interaction between the two
polymers - an interaction induced by the randomness.  For the many
polymer problem (for the $n$-th moment), there is nothing beyond two
polymer interaction.  This has to do with the nature of correlation of
disorder.

\subsubsection{Bound state: two polymer problem and RG}
\label{sec:zrg}
For the second moment, we have a two polymer problem.  The analogy
with quantum mechanics tells us that for $d<2$, any binding potential
can form a bound state but a critical strength is required for a bound
state for $d>2$.  In the polymer language, this means that any small
disorder will change the behaviour of the free (pure) chain for $d<2$
({\it disorder is always relevant}) but for $d>d_c=2$, if $\beta
\Delta <(\beta\Delta)_c$, the chain remains pure-like ({\it disorder
  is irrelevant}).  Actually in higher dimensions ($d>2$) the
delta-function potential needs to be regularized appropriately (e.g.,
by a ``spherical'' well).  Such cases are better treated by
renormalization group (RG) which also helps in making the definitions
of relevance/irrelevance more precise.  We discuss this below \cite{jj}.

In short, the second term (fluctuation in partition function) in the
expansion of Eq. (\ref{eq:2}) cannot be ignored if $d<2$ or if
$\beta\Delta$ is sufficiently large for $d>2$.  This signals a
disorder dominated phase for all disorders in low dimensions or at low
temperatures (strong disorder) in higher dimensions.

\subsubsection{Expansion in potential}
\label{sec:expansion-potential}
We do an expansion in the interaction potential and just look at the
first contributing term.  The full series can of course be treated
exactly.  On averaging, the first order terms drop out, yielding
\begin{equation}
\label{eq:2.1.3a}
\dav{Z^2} - \dav[2]{Z}=
\int {\cal DR}_1\ \int {\cal DR}_2\  e^{-\beta H_{0,1}}\,
e^{-\beta H_{0,2}}  \int  dz\,  \beta^2 \, \Delta\,  \delta(r_1(z)-r_2(z)) + ... .
\end{equation}
This is the first order term if Eq. (\ref{eq:5}) is used.  A
diagrammatic representation is often helpful in book-keeping as shown
in Fig. \ref{fig:ufp}.  Some details of the renormalization group
approach is given in Appendix \ref{app:rg}.

\subsubsection{Reunion}
\label{sec:reunion}
For this two polymer problem, the interaction contributes whenever
there is a meeting or reunion of the two polymers at a site \cite{reuni}.
At the order we are considering, there is only one reunion but this reunion
can take place anywhere along the chain and anywhere in the transverse
direction.  The second order term as shown on the right side of  Fig. \ref{fig:ufp} can be
thought of as two walkers starting from origin have a reunion
anywhere, thereby forming a loop.

\figreu

The probability that two walkers starting from origin would meet at
${\bf r}$ at $z$ is given by $P^2({\bf r},z)$ (Eq. (\ref{eq:1.2})) so
that a reunion anywhere is given by a space integral of this
probability which gives
\begin{equation}
\label{eq:1.5a}
{\cal R}_z  \equiv \int d{\bf r} \, P^2({\bf r},z) = (4\pi z)^{-\Psi}, 
\ {\rm with} \ \Psi=d/2. 
\end{equation}
This exponent $\Psi$ will be called the reunion
exponent.
The
occurrence of a power law is again to be noted.  The eventual
renormalization group approach hinges on this power law behaviour.

It should be noted that the value of the reunion exponent above is
that of simple Gaussian chains.  This need not be the case with
interaction.  For example for repulsive interaction between two
directed polymers, $\Psi=3/2$ in $d=1$ but $ {\cal R}_N \sim N^{-1}
(\ln N)^{-2}$ in $d=2$. The Gaussian behaviour is recovered in $d
>2$ \cite{reuni}.

\subsubsection{Divergences}
\label{sec:divergences}
The contribution in Eq. (\ref{eq:2.1.3a}) at the next one loop order ($O(\Delta^2)$)
as shown in Fig.  \ref{fig:ufp}  involves,
apart from some constants, an integral over the reunion behaviour
given in Eq. (\ref{eq:1.5a}).  This integral in the limit
$N\rightarrow\infty$ is
\begin{equation}
\label{eq:2.1.4}
\int_a^N dz \, \frac{1}{z^{d/2}} \sim {a^{1-d/2}} \quad   {\rm for}\ d>2
\ {\rm but} \sim {N^{1-d/2}}\quad   {\rm for}\ d<2.
\end{equation}
For a finite cut-off, as is usually the case, the integral is finite
for $d>2$, and therefore $\dav{(Z - \dav{Z})^2} \approx
O(\beta^2\Delta)$.  This however is not the case for $d<2$ with $d=2$
as a {\it borderline case}.

The divergence we find for $d<2$ comes from the large $N$ behaviour of
the probability distribution and is therefore ignorant of the details
at the microscopic level. In other words, a lattice model will also
show this divergence in low dimensions.  This forms the basis of a
rigorous analysis done in Ref. \cite{imbrie}, but we pursue a
renormalization group approach here.

\subsubsection{RG flows}
\label{sec:rg-flows}
The problem of divergence we face here is ideal for a renormalization
group approach.  Let us introduce an arbitrary length scale $L$ in the
transverse direction and define a dimensionless ``running'' coupling
constant
\begin{equation}
  \label{eq:1}
  u(L)= (\beta^3 K \Delta) L^{\epsilon}, \quad \epsilon=2-d.
\end{equation}
The purpose behind this length scale is to choose a tunable scale at
which we may probe the system. One may then study the RG flow of the
coupling constant as the scale $L$ is changed.

In the dimensional regularization scheme we adopt here (see Appendix
\ref{app:rg} for details) the divergence seen in Eq.  \ref{eq:2.1.4}
are handled by analytic continuation in $d$.  The problem of
convergence of the integral then appears as singularities at specific
$d$.  One then tries to remove these divergences in $\epsilon$ by
absorbing them in the coupling constants, thereby renormalizing the
coupling.  This in a sense takes care of reunions at small scales to
define the effective coupling on a longer scale.  One then has to
rescale the system to preserve the original length scales.  With this
rescaling, one ends up with a description on a coarser scale with
small scale fluctuation effects getting absorbed in the redefined
parameters.  The fact that the process can be implemented without any
need of additional parameters is linked to renormalizability of the
Hamiltonian.

The effect of renormalization is best expressed by the variation of
the parameters or coupling constants with length scale.  This is
called a flow equation.  For the problem in hand, the  eventual
flow equation 
\begin{equation}
\label{eq:uflow}
L\frac{du}{dL} = (2-d)u + u^2 ,
\end{equation} 
where the first term follows from the definition of $u$ while the
second one is from the loop.  The magnitude of the coefficient of the
$u^2$ term is not very crucial because at this order, this coefficient
can be absorbed in the definition of the $u$ itself.  {\it What
  matters is the sign of the $u^2$ term}.   General cases are
discussed in Appendix \ref{app:flwec}.

\figfp

Since $u$ emanates from the variance of the disorder distribution, it
cannot be negative.  We therefore need to concentrate only on the
$u\geq 0$ part with the initial condition of $u(L=a)=u_0$.  What we
see is that for $d<2$, the flow on the positive axis goes to infinity
indicating a strong disorder phase for any amount of disorder provided
we look at long enough length scale.  An estimate of this length scale
may be obtained from the nature of divergence for a given $u_0$.  An
integration of the flow equation gives $L\sim u_0^{1/(2-d)}$ ($d<2$),
a crossover length beyond which the effect of the disorder is
appreciable. 

For $d>2$, there is a fixed point at $u^*=|\epsilon|$ where
$\epsilon=2-d$.  For $u<u^*$, the disorder strength goes to zero and
one recovers a ``pure''-like behaviour.  This is a \textit{weak
disorder} limit.  But, if $u>u^*$ the disorder is relevant.

Based on the fixed point analysis, we conclude, as already mentioned,
that disorder can be relevant depending on the dimensions we are in
(i.e. the value of $d$) and temperature or strength of disorder.  In
particular, one finds
\begin{enumerate}
\item A disorder-dominated or strong disorder phase for all
  temperatures for $d\le 2$.
\item A disorder dominated or strong disorder phase at low
  temperatures for $d>2$.
\end{enumerate}

For $d>2$, one sees a phase transition by changing the strength of the
disorder or equivalently temperature for a given $\Delta$.  This is an
example of a {\it phase transition induced by disorder} which cannot
exist in a pure case.

It is to be noted that the phase transition (the unstable
$O(|\epsilon|)$ fixed point)
occurs because of the positive $u^2$ term in the flow
equation of Eq. (\ref{eq:1}).  At $d=2$, $u$ is marginal (no $L$
dependence) but renormalization effects lead to an eventual growth.
Such parameters are called marginally relevant.  Any marginally
relevant variable will produce an unstable fixed point, and hence a
phase transition, in dimensions higher than the dimension in which it
is marginal.  A general statement can then be made: Disorder is
expected to produce a new phase transition if it is marginally
relevant at its critical dimension.

The new phase transition (a critical point) is to be characterized by
its own set of exponents.  An important quantity is the length scale
behaviour.  The flow equation around the fixed point for $d>2$ shows
that one may define a diverging ``length-scale'' associated with the
critical point as
\begin{equation}
  \label{eq:6}
 \xi \sim |u-u^*|^{-\zeta} , \ {\rm with}\ \zeta=1/|2-d|.
\end{equation}
In the critical dimension $d=2$, there are log corrections.  In the
weak disorder phase where the disorder is irrelevant, $\dav{\ln Z}
\approx \ln \dav{Z}$, and therefore one may put a bound on the
transition temperature $T_c$ for a lattice model as $T_c\geq T_A$ as
defined below Eq. (\ref{eq:2.1.5}).

Attempts were made to determine $\zeta$ by numerical methods and
verify the RG prediction.  However, the results from specific heat
 \cite{spheat} and size calculations \cite{kim} agree neither with each
other nor with the RG result of Eq. (\ref{eq:6}).  This remains an open
problem.

We come back to the strong disorder phase in Sec. \ref{sec:strong}.

\subsection{Bethe ansatz and $\theta$}
For the directed polymer problem, a mapping to a quantum problem helps
in the evaluation of $\dav{Z^n}$ at least in $d=1$.  For a Gaussian
distributed, delta-correlated disorder, $\dav{Z^{n}}$ corresponds to
the partition function of an $n$-polymer system with the Hamiltonian
given by Eq. (\ref{eq:5}).  Noting the similarity with the quantum
Hamiltonian with $z$ playing the role of imaginary time, finding
$N^{-1} \ln\dav{Z^n}$ for $N\rightarrow\infty$ is equivalent to
finding the ground state energy $E$ of a quantum system of $n$
particles.  This problem can be solved exactly only in one dimension
($d=1$) using the Bethe ansatz \cite{Bethe,kard}.  This gives the
ground state energy as
\begin{equation}
  \label{eq:8}
 E= - K(n - n^3 ) \ {\rm in}\  d=1,
\end{equation}
which gives 
\begin{equation}
  \label{eq:11}
  \theta=\frac{1}{3},
\end{equation}
from Eq. (\ref{eq:rep.3}) As we shall see below, this implies
$\nu=2/3$ so that the polymer has swollen far beyond the random walk
or Gaussian behaviour.  What looks surprising in this approach is that
there is no ``variance'' (2nd cumulant) contribution.  It is just not
possible to have a probability distribution whose variance vanishes
identically.  This is a conspiracy of the $N\rightarrow \infty$ limit
inherent in the quantum mapping and the value of the exponent $\theta$
that suppressed the second cumulant contribution (see
Eq. (\ref{eq:rep.3})).

\subsubsection{Flory approach}
Using the quantum analogy, we may try to estimate the ground state
energy in a simple minded dimensionally correct calculation based on
the assumption of only one length scale.  Such an approach generally
goes by the generic name of ``Flory approach''.  The elastic energy is
like the kinetic energy of quantum particles which try to delocalize
the polymers (random walk) while the attractive potential tries to
keep the polymer together.  For $n$ polymers there are $n(n-1)/2$
interactions.  We take the large $n$ limit so that if the particles
are bound in a region of size $R$, the energy is (using dimensionally
correct form with $R$ as the only length scale)
\begin{equation}
\label{eq:beth.4}
E= \frac{n}{R^2} - \frac{n^2\Delta}{R}
\end{equation}  
which on minimization gives $E\sim n^3$ consistent with the Bethe
ansatz solution.  At this point we see the problem of the replica
approach if the limit is taken too soon.  Since our interest is
eventually in $n \rightarrow 0$, we could have used in this argument
the linear term of the combinatorics.  That would have made energy
``extensive'' with respect to the number of particle and replaced the
disorder-induced attraction by a repulsion (note the negative sign).
The end result would however have no $n^3$ dependence.  This is a real
danger and any replica calculation has to watch out of these pitfalls.
Quite strangely we see that the correct answer comes by taking
$n\rightarrow\infty$ first and then $n\rightarrow 0$ or, probably
better to say, by staying along the ``attractive part'' of the
interaction only.

\subsubsection{Confinement energy}
\label{sec:confine}
Suppose we confine the polymer in a tube of diameter $D$. This is like
the localization length argument used to justify the energy in the
quantum formulation.  The polymer in the random medium won't feel the
wall until its size is comparable to that, $D\sim N_0^{\nu}$ which
gives the length at which the polymer feels the wall.  Elastic energy
of a blob is $D^2/N_0$.  But because of the tube, the polymer will be
stretched in the tube direction.  One may then consider the polymer as
consisting of free $N/N_0$ blobs aligning with the force, so that the
energy is
\begin{equation}
\frac{N}{N_0} \ \frac{D^2}{N_0} = N \ 
\left(\frac{D}{N_0}\right )^{^{\scriptstyle 2}}
= N\ \frac{1}{D^{2(1-\nu)/\nu}}.
\end{equation}
This gives the known form $1/D^2$ used in the quantum analogy,
Eq. (\ref{eq:beth.4}), ( and consistent with dimensional analysis) but
for $\nu=2/3$, this gives $1/D$.

A cross-check of this comes from the energy of a blob. Each blob has
the fluctuation energy $N_0^{\theta}$ and so free energy per unit
length $F/N\sim N_0^{\theta}/N_0\sim D^{-2(1-\nu)/\nu}$.

\section{Analysis of free energy: specialty of directed polymer}
\label{sec:can-we-handle}
Another unique feature of this directed polymer problem is that there
is a way to study the average free energy and implement RG directly
for the free energy bypassing the $n\rightarrow 0$ problem of the
replica analysis completely, giving an independent way of checking the
results of replica approach.
 
\subsection{Free energy and the KPZ equation }
For a polymer, the partition function satisfies a diffusion or
Schrodinger-like equation. This equation can be transformed to an
equation for the free energy $F({\bf r},z)=-T \ln Z({\bf r},z)$.  This
is the free energy of a polymer whose end point at $z$ is fixed at
${\bf r}$.  To maintain the distance fixed at ${\bf r}$ a force is
required which is given by ${\bf g}=-\nabla F$.  If we want to
increase the length of a polymer by one unit, we need to release the
constraint at the previous layer (think of a lattice). The change in
free energy would then depend on the force at that point, and of
course the random energy at the new occupied site. The change
$\partial F({\bf r})/\partial z$ being a scalar can then depend only
on the two scalars $\nabla \cdot {\bf g}$ and  $g^2$.  A direct derivation
of the differential equation for the free energy shows that these are
the three terms required.  The differential equation, now known as the
Kardar-Parisi-Zhang equation \cite{kard2}, is
\begin{equation}
\label{eq:kpz}
\frac{\partial F}{\partial z} = \frac{T}{2K} \nabla^2 F - \frac{1}{2K}
(\nabla F)^2 + \eta({\bf r},z). 
\end{equation}
If we can solve this \textit{exact} equation and average over the
random energy $\eta$, we get all the results we want.

One may also write down the equation for the force in this ``fixed
distance'' ensemble as
\begin{equation}
  \label{eq:12}
\frac{\partial {\bf g}}{\partial z} =  \frac{T}{2K} \nabla^2  {\bf g}
     - \frac{1}{2K} {\bf g}\cdot\nabla {\bf g} + \nabla \eta({\bf  r},z).
\end{equation}
This equation is known as the Burgers equation.

\subsection{ Free energy of extension: pure like}
\label{sec:free-energy-extens}
We want to know the free energy cost in pulling a polymer of length
$N$ from origin (where the other end is fixed) to a position ${\bf
r}$.  For the pure case, the free energy follows from
Eq. (\ref{eq:1.2}) (with $K$ inserted) as
\begin{equation}
\label{eq:1.4}
F({\bf r},N)- F({\bf 0},N) = \frac{d}{2} \ K \ \frac{ r^2}{N}. \ ({\rm
  pure})
\end{equation}
The probability distribution for the free energy can be obtained by
choosing ${\bf g}= {Kd \bf r_N}/N$ in Eq. (\ref{eq:2.3.1}) as.
\begin{equation}
  \label{eq:13}
 P(F({\bf r}_N)) = P(F(0) + \frac{Kd r_N^2}{2N}) .
\end{equation}
One then gets the surprising result,
\begin{equation}
  \label{eq:14}
\dav{F({\bf r},N)- F({\bf 0},N)} = \frac{d}{2} \ K \ \frac{ r^2}{N}.\ ({\rm
 disorder})
\end{equation}
Therefore, on the average the stretching of a chain is pure-like
(elastic) with the same elastic constant though the fluctuation is
anomalous ($\theta \neq 0$).  This is analogous to the pure-like
result for the correlation function, Eq. (\ref{eq:2.3.4}). These
results have a far reaching consequence that in a renormalization
group procedure, the elastic constant must remain an invariant
(independent of length scale).  As we shall see, this invariance
condition puts a constraint on $\nu$ and $\theta$, making only one
independent.

\subsection{RG of the KPZ equation}
To analyze the nonlinear KPZ equation, an RG procedure may be adopted.
This RG is based on treating the nonlinear term in an iterative manner
by starting from the linear equation.  This is a bit unusual because
here we are not starting with a ``Gaussian'' polymer problem, rather,
a formal linear equation\cite{ew} that \textit{does not} necessarily represent
a polymer.  Leaving aside such peculiarity, one may implement the
coarse-graining of RG to see how the couplings change with length
scale.

\subsubsection{Scale transformation and an important relation}
\label{sec:scale-transf-an}
Under a scale transformation $x\rightarrow bx, z\rightarrow b^{1/\nu}
z$, and $F \rightarrow b^{\theta/\nu}F$, the randomness transforms
like $\Delta \rightarrow b^{-d-(1/\nu)} \Delta$.  This transformation
done on Eq. (\ref{eq:14}) shows that for $K$ to be an invariant (no
$b$-dependence) we must have,
\begin{equation}
  \label{eq:15}
\theta+1 = 2\nu.
\end{equation}
This is trivially valid for the Gaussian pure polymer problem but
gives a relation between the free energy
fluctuation
and the size of the
polymer.  This is borne out by the intuitive picture we develop below.
This relation gives the size exponent $\nu=2/3$ in $d=1$.

The equation in terms of the transformed variables is then
\begin{equation}
\frac{\partial F}{\partial z} = \frac{T}{2K}b^{(1-2\nu)/\nu}\, \nabla^2
F - \frac{1}{2K} b^{(\theta-2\nu+1)/\nu} \,(\nabla F)^2 +
 b^{(1-d\nu-2\theta)/\nu}\, \eta({\bf r},z)
\end{equation}
The $b$-dependent factors can be absorbed to define new parameters,
except for $K$.

The temperature however gets renormalized as $T \rightarrow
T\,b^{(1-2\nu)/\nu}$.  Its flow is described by the flow
equation
\begin{equation}
\label{eq:1.2.2}
L\frac{\partial T}{\partial L} = \frac{1-2\nu }{\nu}\, T  \ ({\rm to \
  leading \ order})
\end{equation}
For $\nu>1/2$, $T(L) \rightarrow 0$.  The disorder dominated phase is
therefore equivalent to a zero temperature problem.  In other words,
the fluctuation in the ground state energy and ground state
configurations dominate the behaviour at low temperatures in
situations with $\nu>1/2$.  It is this renormalization that was
missing in Sec \ref{sec:zrg}.

\subsection{RG flow equation}
The nonlinearity (or $g^2$) contributes further to the renormalization
of the temperature through the appropriate dimensionless variable
$u=(K\Delta/T^3) L^{2-d}$ (same as in Eq. (\ref{eq:1}).  As in the
previous section,  the important
flow equation 
is for this
parameter $u$ (upto constant factors).  The renormalization of
temperature acquires subtle $d$-dependence that introduces a new
element in the flow equation.  Some details may be found  in Appendix
\ref{app:kpz}. We quote the flow equation here as
\begin{equation}
  \label{eq:21}
L\frac{du}{dL} = (2-d)\, u + \frac{2d-3}{2d}\,  u^2.
\end{equation}
  General cases of such a flow
equation  are discussed in Appendix \ref{app:flwec}.

One immediately sees a major difference with the flow equation Eq.
(\ref{eq:uflow}) for $d=1$.  Because of the change of sign of the
quadratic term in Eq. (\ref{eq:21}), there is now a fixed point for
$d=1$ as shown in Fig. \ref{fig:fp}(b).  This flow equation does not
behave properly in a range $d\in [1.5,2)$ but that is more of a
problem of implementation of RG than directed polymer {\it per se},
and so, may be ignored here.  Note also that no extra information can
be obtained from Eq. (\ref{eq:21}) for $d\ge 2$ other than what we have
obtained so far in Sec. \ref{sec:disorder-relevant}, namely the
existence of a critical point.  

This approach however has the advantage of getting the renormalization
of temperature by $u$. Some details of this RG is given in Appendix
\ref{app:kpz}. Eq. (\ref{eq:1.2.2}) then gets modified to
\begin{equation}
\label{eq:1.2.3}
 L\frac{\partial T}{\partial L} = \left (\frac{1-2\nu }{\nu} + \frac{2-d}{4d} u\right ) T. 
\end{equation}
If we now demand scale invariance at a fixed point of $u$, we can
determine the exponent $\nu$.  At d=1, the stable fixed point $u^*=2$
from Eq. (\ref{eq:21}) then gives the exact
exponents at $d=1$
\begin{equation}
\label{eq:beth.3}
\nu=\frac{2}{3}, \theta=\frac{1}{3}, 
\end{equation}
in agreement with the Bethe ansatz results mentioned above. To get
$\theta$ the exponent relation Eq. (\ref{eq:15}) (from invariance of $K$)
has been used.  The RG results for $d=1$ are expected to be exact.

\subsection{Critical point for $d>2$}
For $d>2$, the unstable fixed point is $u^*=O(|\epsilon|)$, 
($d=2-\epsilon$).  This gives at the critical point $\nu=1/2
+O(\epsilon^2)$ indicating the possibility of 
\begin{equation}
\label{eq:1.2.4}
\nu=1/2, \ {\rm and} \ \theta=0 
\end{equation}
to be exact.

One may argue \cite{doty} for $\theta=0$  in the following way.  At $T=T_c$, thermal
fluctuation enables the polymer to get out of the trap set by the
random potential (``ground state'').  Just above the critical point, on a scale determined
by the correlation length of the critical point, the random
potential scaling is set by $b^{\theta}$ with the value of $\theta$ at
$T_c$. As $T\rightarrow T_c+$, the length scale diverges and therefore the
relevant energy scale would also grow with the same exponent.  However
a critical point implies the energy scale to be $O(T_c)$ which is
finite.  These can be reconciled if $\theta=0$ at the critical
point.  One then gets $\nu=1/2$.  Though this is the same as that of a
Gaussian polymer, we shall see later that the polymer has extra
non-Gaussian features.

\subsection{Strong disorder phase for $d\geq2$}
\label{sec:strong}
Unlike the strong disorder phase at d=1, the absence of any fixed
point for the strong disorder phase for $d>2$ in this approach forbids
quantitative results about the phase itself.  In addition, the
behaviour of the strong disorder phase in $d=1$ can be obtained by
various methods.  Unfortunately, there are few concrete results in
higher dimensions $d\geq 2$.  Most reliable values of the exponents
come from various numerical approaches based on the KPZ equation.  
A recent estimate for $d=2$ is \cite{giada}
\begin{equation}
  \label{eq:63}
 1/\nu=1.67\pm 0.025,\ {\rm and}\ \theta = 0.229\pm0.05.
\end{equation}

Numerical studies indicate that $\nu$ decreases as $d$ increases.  A
question arises about the existence of an upper critical dimension
$d=d_{\rm UCD}$ such that $\nu =1/2$ for $d>d_{\rm UCD}$.  For example,
higher loop contributions in the RG of Sec.  \ref{sec:rg-flows} show
singularities at $d=4$ which could indicate $d=4$ as another critical
dimension. Various analytical approaches \cite{hhz,lass,jkb} suggest
$d_{\rm UCD}=4$, or even nonintegral dimension \cite{blum}.  But
numerical simulations and other arguments \cite{marinari,castel,HF}
suggest $d_{\rm UCD}=\infty$.  This issue is yet to be resolved.

The fact that the size exponent $\nu$ (often called wandering
exponent)
is different from $1/2$ has
important implications in various applications, especially for flux
lines in superconductors.  For example, confinement of a flux line in
presence of many other flux lines would lead to a steric repulsion\cite{nat}
(similar to the confinement energy in Sec. \ref{sec:confine}) and the
interaction of the vortices may lead to an attractive fluctuation
induced (van der Waals type) interaction\cite{sm2}.

\section{Overlap}
\label{sec:over}
In a replica approach, 
occupancy of different ground
states may be achieved by ``replica symmetry breaking'' (i.e. various
replicas occupying various states) but the difficulty arises from the
$n\rightarrow0$ limit.  In the case of directed polymer, we have
argued that the degenerate states occur only rarely and therefore the
effect of ``replica symmetry breaking'' 
if any has to be very small.  This is why the Bethe ansatz
gave correct results without invoking replica symmetry breaking.

The method to compute the overlap was developed by Mukherji \cite{sm}.
By introducing a repulsive potential $v\int dz \delta({\bf
r}_1(z)-{\bf r}_2(z))$ for the two polymers in the same random medium,
\begin{eqnarray}
{\sf H}&=&\frac{1}{2} \int_{0}^{N} dz \
\left[\left(\frac{\partial
{\bf{r}}_1(z)}{\partial z}\right)^{^{\scriptstyle 2}}+
\left(\frac{\partial{\bf{r}}_2(z)}{\partial
z}\right)^{^{\scriptstyle 2}}\right]+ \int_0^N dz\  \eta({\bf r}_1(z),z)\nonumber\\
&&\qquad \qquad + \int_0^N dz\  \eta({\bf r}_2(z),z) +
\int_{0}^{N} dz \ v_0\ \delta{\bf(}{\bf{r}}_{12}(z){\bf )}.
\end{eqnarray}
The free energy $F({\bf r}_1(z),{\bf r}_2(z),z,v)$ satisfies
a modified KPZ type equation
\begin{equation}
  \label{eq:20}
\frac{\partial F}{\partial z} = \sum_i \left (\frac{T}{2K} 
\nabla^2_i F - \frac{1}{2K}
(\nabla_i F)^2 + \eta({\bf r}_i,z)\right)+ v\delta({\bf r}_1-{\bf r}_2).   
\end{equation}
which can be studied by RG.  The mutual interaction has no effect on
the single chain behaviour but the interaction gets renormalized.  The
flow equation for the dimensionless parameter $u$ of Eq. (\ref{eq:21})
remains the same.  The exponent relation of Eq. (\ref{eq:15}) also
remains valid.  The interaction gets renormalized as
\begin{equation}
  \label{eq:22}
  L\frac{\partial v}{\partial L}= \left(\frac{1-\theta}{\nu} -d +
    \frac{u}{2}\right) v ,
\end{equation}
where $v$ is in a dimensionless form.  For the pure problem
($\theta=0,\nu=1/2$) this reduces to the expected flow equation of Eq.
(\ref{eq:uflow}) for repulsive interaction ($ u \rightarrow -u$). For
overlap one needs only the first order term because we need
$v\rightarrow 0$.
 
The overlap can be written in a polymer-type scaling form $q
=N^{\Sigma}\ {\sf f}(v N^{-\phi\nu})$, where $\Sigma=\theta- \phi\nu
-1$.  The above RG equation for $v$ shows that the exponent $\Sigma
=0$ at the stable fixed point 
for $u$ of Fig \ref{fig:fp}(b) at $d=1$.  However,
$\Sigma< 0$ at the transition point for $d>2$.  This means that the
overlap vanishes at the transition point from the strong disorder side
as $q\sim |T-T_c|^{|\Sigma|\zeta}$. 

This approach to overlap can be extended to $m$-chain overlaps also,
which show a nonlinear dependence on $m$ at the transition
point \cite{smsmb}. This suggests that eventhough the size
exponent
is $\nu=1/2$ Gaussian like, there is
more intricate structure than the pure Gaussian chain.  Overlaps of
directed polymers on trees have been considered in Ref.  \cite{der}.  A
case of cross-correlation of randomness (each polymer seeing a
different noise) has been considered by Basu in Ref.  \cite{abhik}.

\section{ Unzipping: pure case}
\label{sec:unz_pure}
Unzipping was first considered in the context of DNA \cite{smb_mt}.
However the same ideas play a role here.  Let us consider a pure case
of a directed polymer with one end fixed at origin and with an
attractive interaction with a line at ${\bf r}(z)=0$ (instead of being
in a random medium).  The Hamiltonian for a delta-function interaction
can be written as
\begin{equation}
H=\frac{d}{2} K \int_0^N dz \left( \frac{\partial {\bf r}}{\partial
z}\right ) ^{^2} -  v \int_0^N dz \ \delta({\bf r}(z)) -  
{\bf g}\cdot \int_0^N\  \ dz\  \frac{\partial {\bf r}}{\partial z},
\end{equation}
very similar in form with Eq. (\ref{eq:2.3.1}) except here we have an
attractive interaction instead of a random medium.   

For the zero force case, there is a critical unbinding transition at
$v=v_c$.  For $d\le 2$, $v_c=0$.  The pulling force would like to
align the polymer in the direction of the force while the interaction
would like to keep the polymer attached to the rod.  At zero
temperature the unzipping transition
takes place at a force where the binding energy is compensated by the force
term.  Upto a geometric factor $a$, this is given by $Nv=a g N$.  At
nonzero temperature, the entropic effects are to be taken into
account, which may be done by using the quantum analogy.

The problem can be mapped on to a quantum Hamiltonian, albeit
non-hermitian, for a particle of co-ordinate ${\bf r}$
\begin{equation}
\label{eq:qham}
H_q({\bf g}) = \frac{1}{2} ({\bf p} + i {\bf g})^2 + V({\bf r}),
\end{equation}
in units of $\hbar (\equiv k_B T) =1$ and mass $=1$, with ${\bf p}$ as
momentum.  For long chains $N\rightarrow \infty$ the free energy is
the ground-state energy of this non-hermitian Hamiltonian. A phase
transition takes place whenever the ground state is degenerate.  The
analysis done in Ref.  \cite{smb_mt} shows that if the ground state
energy ( i.e. the binding energy of the polymer per unit length) is
$E_0$, then the critical force is given by
\begin{equation}
  \label{eq:2pr}
  g_c=2\sqrt{E_0} \sim  |v-v_c|^{1/|2-d|} .
\end{equation}
where the $v$-dependences of $E_0$ close to $v_c$, for general $d$, is
used.  In fact if the bound state has extensive entropy, then there is
a possibility of a re-entrance at low temperatures (see
Sec. \ref{sec:unzip}.
This however is not possible in this case in hand.

\section{Nature of ground states and excitations}
\label{sec:intuitive-picture}

Powered by the quantitative estimates of the free energy fluctuation
and size exponents, we now try to generate a physical picture.

\subsection{Rare events}
\label{sec:intu-pict-rare}
We have seen that there is a low temperature region (in lower
dimensions for all $T$) where randomness results in a new phase but
the response to an unzipping force
is the same as for the pure system.  For the pure case as 
$N\rightarrow\infty$ the
width of $P({\bf r}_N,N)$ increases.  Hence the increase of $C_T$ with
$N$.  With randomness, for $T\rightarrow 0$ we need to look for the
minimum energy path.  Let us suppose that there is a unique ground
state, i.e.  $E({\bf r}_N)$ or $F({\bf r}_N)$ is a minimum for a
particular path.  This tells us that as the temperature is changed,
$T$ still low, the polymer explores the nearby region so that the
probability distribution gains some width which is determined by the
thermal length.  Susceptibility would be the width of the distribution
and this is independent of $N$.  This cannot satisfy the relation
given by Eq. (\ref{eq:15}). If we invoke the the unzipping argument, then we
need to exert a force exceeding the critical force to take the polymer
out of the bound state and so the response to a small force
($g\rightarrow 0$) would be insignificant.  The situation will not be
any better even on averaging over the random samples if every sample
has a unique ground state.

However, it may happen that most of the samples have unique
ground-states but once in a while (rare samples) there is more than
one ground state which happens to be far away from each other.
Suppose there are such rare samples, whose probabilities decay as
$N^{-\kappa}$, where the paths are separated by $N^{\nu}$, then the
contribution to the fluctuation from these samples would be $N^{2\nu
  -\kappa}$.  In case $\kappa=2\nu -1$, we get back the exact result.
The relation of Eq. (\ref{eq:15}) tells us $\kappa=\theta$.  The rare
events control the free energy fluctuation.  From the unzipping point
of view, the threshold in such rare cases is at zero force because a
small force can take a polymer from one ground state to another one,
gaining energy in the process.  Following Ref.  \cite{mez}, one may
argue that the gain in energy from the force should be similar to the
energy fluctuation.  Assuming a scaling of the force, $g\sim {\hat
  g}N^{\sigma}$, then ${\hat g}N^{\sigma}N^{\nu} \sim N^{\theta}$
which gives $\sigma=\theta-\nu=1/3$ in $d=1$. This argument implies
that if the average stretching is proportional to $g$ and to $N$,
i.e., $\dav{<r>}\sim gN$, then one should get a linear plot if
$\dav{<r>}/N^{\nu}$ is plotted against $gN^{\nu-\theta}$.  The
surprising feature is the sample dependence of such a plot.  For a
directed polymer, these quantities can be obtained by a transfer
matrix calculation which is exact for a given sample and finite $N$.
Some details on the transfer matrix approach are given in Appendix
\ref{app:transf}.  As shown in Fig. \ref{fig:step}, one sees steps
with an overall linear dependence.  The susceptibility within a flat
step is zero as seen in the plot of the fluctuations.

\figsteps

What we see here is that though the average behaviour is the same as
that of the pure system, the underlying phenomenon is completely
different; the average thermal response is determined by the rare
samples that have widely separated degenerate ground state and the
probability of such states also decays as a power law of the chain
length.  This picture also shows the ensemble dependence.  What we
have discussed is the fixed force ensemble.  In the fixed distance
ensemble, 
if we keep the end point
fixed at ${\bf r}$ and try to determine the force to maintain it at
that point, then by definition, the force comes from a small
displacement around ${\bf r}$.  Such small displacements will never
lead to the big jumps that ultimately contribute to the average
susceptibility.  This difference in behaviour in the two ensembles is
one of the important features of quenched randomness.

In a given sample elastic energy $\sim {r^2}/N \sim N^{2\nu -1}$.  The
pinning energy would also grow with length say as
$N^{\tilde{\theta}}$.  We see, $\tilde{\theta}=2\nu -1=\theta$.  One
way to say this is that the sample to sample fluctuation and the
energy scale for a given sample are the same.

\figriv

These results can now be combined for an image of the minimum energy
paths.  If the end points at $z=N$ are separated by $r \ll N^{\nu}$,
the paths remain separated (each path exploring an independent
disordered region) until they join at $\Delta z\sim r^{1/\nu}$, after
which they follow the same path.  If the end points are separated by a
distance $r \gg N^{\nu}$, the two paths explore independent regions
and they need not meet.  This picture (Fig.\ref{fig:river}) is often
alluded to as the river-basin network.

\subsection{Probability distribution}
For a pure polymer, the probability distribution of the end point is
Gaussian but it need not be so for the disordered case.  One way to
explore the probability distribution is to study the response of the
polymer as we take it out of its optimal or average position, e.g. by
applying a force.  In a previous section we used the fixed distance
ensemble where the end point was kept fixed and we looked at the force
$g$ required to maintain that distance (see Eq. (\ref{eq:12})). Here
we consider the conjugate fixed force ensemble.

\subsubsection{Response to a force}
\label{sec:response-force}
Let us apply a force that tries to pull the end of the polymer beyond
the equilibrium value $r \sim R_0$.  In equilibrium, the average size
$R$ or extension by the force can be expressed in a scaling form
\begin{equation}
\label{eq:1.6}
R= R_0 \ {\sf f}\!\left(\frac{gR_0}{ T}\right).
\end{equation}
This is because for zero force one should get back the unperturbed
size while the force term may enter only in a dimensionless form in
the above equation where the quantities available are the size $R_0$
and the thermal energy.  For small $g$, linearity in $g$ is
expected. This requirement gives
\begin{equation}
\label{eq:1.7}
R= R_0 \frac{R_0}{T} g \quad (k_B=1),
\end{equation}
$R$ is not proportional to $N$ if $\nu \neq 1/2$ ($R_0 \sim N^{\nu}$).
The polymer acts as a spring with $T/R_0^2$ as the effective spring
constant.

\subsubsection{Scaling approach}
\label{sec:scaling-approach}
Let us try to develop a physical picture and the corresponding
algebraic description (called a scaling theory).  The polymer in
absence of any force has some shape of characteristic size $R_0$.  The
force stretches it in a way that it breaks up into blobs
of size $\xi_g=T/g$.  For size $< \xi_g$ the polymer looks
like a chain without any force and these blobs, connected linearly by
geometry, act as a ``new'' polymer to respond to the force by aligning
along it.  We now have two scales $R_0$ and $\xi_g = T/g$.  A
dimensionless form is then
\begin{equation}
\label{eq:1.8}
R= R_0\  {\sf f}\!\left(\frac{R_0}{\xi_g}\right) \sim  \frac{R_0^2}{T} \ g
\end{equation}
Now each blob is of length $N_g$ so that $\xi_g= N_g^{\nu}$ and there
are $N/N_g$ blobs.  We therefore expect 
\begin{equation}
\label{eq:1.8a}
R= \frac{N}{N_g} \xi_g = N \xi_g^{1-1/\nu} = N \left
( \frac{g}{T}\right)^{^{(1-\nu)/\nu}}
\end{equation}
This gives a susceptibility $\chi = \partial R/ \partial g \sim
g^{(1-2\nu)/\nu}$ .

\subsubsection{Probability Distribution }
\label{sec:so-prob-distr}
Let us try to get the susceptibility of Eq. (\ref{eq:1.8a}) in another
way.  Let us assume that the probability distribution for large $R$ is
\begin{equation}
  \label{eq:16}
  P(r) \sim \exp\left(- ({r}/{R_0})^{\delta}\right).
\end{equation}
The entropy is given by $S(r) = -\ln P$.  The free energy in presence
of a force which stretches the polymer to the tail region is given by
\begin{equation}
  \label{eq:17}
 F=T({r}/{R_0})^{\delta} - g r. 
\end{equation}
This on minimization gives 
\begin{equation}
  \label{eq:61}
 g= \frac{T}{R_0}\  \left(\frac{r}{R_0}\right)^{\delta-1}. 
\end{equation}
By equating this form with Eq. (\ref{eq:1.8a}), we get
\begin{equation}
  \label{eq:62}
\delta=\frac{1}{1-\nu}  
\end{equation}
and
\begin{equation}
\label{eq:1.9}
P(R) \sim \exp\left[-\left(\frac{r}{R_0}\right)^{{1/(1-\nu)}}\right].
\end{equation}
For $\nu=1/2$ we do get back the Gaussian distribution. 


The above analysis, done routinely for polymers, relies on the fact
that there is only one length scale in the problem, namely, the size
of the polymer.  If we are entitled to do the same for the disorder
problem, namely only one scale, $R_0 \sim N^{\nu}$, matters, then the
blob picture goes through in toto.  The chain breaks up into ``blobs''
and the blobs align as dictated by the force.  Each blob is
independent and the polymer inside a blob is exploring its environment
like a directed polymer pinned at one end.  The probability
distribution is therefore given by Eq. (\ref{eq:1.9}) which for $d=1$ is
\begin{equation}
  \label{eq:18}
 P(r) \sim \exp(-|r|^3/N^2)  \quad (d=1).
\end{equation}
If we use the relation $\Delta F\equiv F(x,N)-F(0,N)\sim x^2/N$, then
the above probability distribution can be mapped to the distribution
of the free energy as
\begin{equation}
  \label{eq:19}
  P(F)  \sim
  \exp\left[-\left(\frac{|\Delta
        F|}{N^{\theta}}\right)^{1/2(1-\nu)}\right]
\sim \exp \left(- \frac{|\Delta F|^{3/2}}{N^{1/3}}\right ) \quad (d=1).
\end{equation}

This has been tested numerically \cite{hhz}.  See,
e.g. Ref.  \cite{prae} for more recent work.

\section {Random interaction - RANI model}
\label{sec:rani}
So far we have been considering the problem of random medium.  A
different situation arises if there is randomness in the interaction
of polymers. This is the RANI model \cite{sm93:1,sm95:2}.
  Consider the problem of two directed polymers
interacting with a short range interaction as in Eq. (\ref{eq:20}) but
take $v$ to be random.  Such problems are of interest, especially in
the context of DNA where the base sequence provides heterogeneity
along the chain.  In this DNA context, the randomness can  be
taken to be dependent only on the $z$ coordinate and not on others
like the transverse position ${\bf r}$.  It can be written as
\begin{equation}
\label{rani:1}
H_{\rm int}=\int_0^N  dz\  v_0\ [1+b(z)]\ \delta({\bf r}_1(z)-{\bf r}_2(z)),
\end{equation}
where the randomness in introduced through $b(z)$.  We take
uncorrelated disorder with a Gaussian distribution
\begin{equation}
\label{rani:2}
\dav{b(z)}=0, \dav{b(z_1)b(z_2)}= \Delta \delta(z_1-z_2).
\end{equation}
This would correspond to uncorrelated base sequence of a DNA problem.
The full Hamiltonian can be written as 
\begin{equation}
{\sf H}=\frac{1}{2} \int_{0}^{N} dz \
\left[\left(\frac{\partial
{\bf{r}}_1(z)}{\partial z}\right)^{^{\scriptstyle 2}}+
\left(\frac{\partial{\bf{r}}_2(z)}{\partial
z}\right)^{^{\scriptstyle 2}}\right]+
\int_{0}^{N} dz \ v_0\ [1+b(z)]\ V{\bf
(}{\bf{r}}_{12}(z){\bf )}.
\label{eq:h}
\end{equation}
where ${\bf r}_i(z)$ is the $d$-dimensional position vector
of a point of chain $i$ at a contour length $z$, and ${\bf
r}_{12}(z)={\bf r}_1(z)-{\bf r}_2(z)$.  Though written for  a general
potential $V({\bf r})$, we shall consider only short-range interaction
that, when convenient, can be replaced by a $\delta$-function.

\subsection{Annealed case: two chain}
As expected, the average partition function (annealed averaging) would
correspond to a pure-type problem.  This however is not the case
always as we see in Sec. \ref{sec:ann3} for more than two chains.
  
A straightforward
averaging of $Z=\int Dr_1\ Dr_2\ \exp (-{\sf H})$ using the probability distribution of
Eq.~(\ref{rani:2}) defines an effective Hamiltonian ${\cal
H}_{{\rm eff}}$ such that
\begin{equation}
\dav{Z} =\int Dr_1\ Dr_2\ \exp (-{\cal H}_{{\rm eff}}),\label{eq:za}
\end{equation}
and it is given by
\begin{equation}
{\cal H}_{{\rm eff}}=\frac{1}{2} \int_{0}^{N} dz \
\sum_{i=1}^{2}\left(\frac{\partial
{\bf{r}}_i(z)}{\partial z}\right)^{^{\scriptstyle 2}} +
v_0\int_{0}^{N} dz\ V{\bf (}{\bf r}_{12}(z){\bf )} -
\frac{v_0^2\Delta}{2}\int_{0}^{N} dz
\ V^2{\bf (}{\bf r}_{12}(z){\bf )}. \label{eq:h1}
\end{equation}
It appears from the above expression of the effective
Hamiltonian that an attraction is generated between the two
chains.  
Now, since any short range potential under renormalization
maps onto a $\delta$ function potential, we can take the
``minimal" effective Hamiltonian for $\langle Z\rangle$ as
\begin{equation}
{\cal H}_{2}=\frac{1}{2} \int_{0}^{N} dz \
\left[\left(\frac{\partial
{\bf{r}}_1(z)}{\partial z}\right)^{^{\scriptstyle 2}}+
\left(\frac{\partial{\bf{r}}_2(z)}{\partial
z}\right)^{^{\scriptstyle 2}}\right]+
{\bar{v}}_0\int_{0}^{N} dz\ \delta{\bf (}{\bf
r}_{12}(z){\bf )}.
\label{eq:hm2}
\end{equation}
where ${\bar{v}}_0$ is the reduced coupling constant which
takes care of the attraction described earlier. We believe
that the large length scale properties as described by
Eq.~(\ref{eq:hm2}) is same as that of Eq.~(\ref{eq:h1}).  If
necessary, we can restrict the strength of the disorder so
that $\bar{v}_0$, which represents the effective coupling
between the two chains, is positive (i.e.  repulsive
interaction).  Now the problem reduces to a relatively
simple situation where the two chains interact with a pure
$\delta$-function interaction with a reduced coupling
constant ${\bar{v}}_0$.  The solution of this pure problem
is known \cite{jj} as discussed in the Appendix \ref{app:rg}.

\figrani

\subsection{Marginal Relevance of disorder}
When we consider the second cumulant of the partition function, we
require four chains.  On averaging over the disorder, a new
``inter-replica'' interaction is generated that couples the original
chains with the replica,
\begin{equation}
\label{rani:3}
H_{\rm rep}= - {\bar{r}_0} \int_0^N dz\ \delta({\bf r}_{12}(z))
\delta({\bf r}_{34}(z)),
\end{equation}
with ${\bar{r}_0}=v_0^2\Delta$ and $3,4$ representing the two new
chains, and ${\bf r}_{ij}(z)={\bf r}_i -{\bf r}_j$.  This is a special
four chain interaction in the sense that this interaction favours a
contact for chains 3 and 4 at $z$ if chains 1 and 2 also enjoy a
contact there, though not necessarily at the same point in the
transverse space (see Fig. \ref{fig:rani}).  In addition to this four
chain interaction, the disorder also creates an effective two body
attraction that changes the starting or bare interaction. So far as
the effective two body interaction is concerned its effect on the long
length scale is given by the RG flows of Eq. (\ref{eq:uflow}) except
that $u$ can now be negative also. Assuming that we are at the
critical point of this two body interaction, e.g. for low dimensions,
at $u^*=0$, the effect of the disorder can be obtained from the RG
flow of ${\bar{r}_0}$.  Defining the dimensionless coupling constant
$r_0={\bar{r}_0}L^{2\epsilon'}$ where $\epsilon'=1-d$ with $r(L)$ as
the dimensionless running coupling constant, the RG equation is given
by \cite{sm93:1,sm95:2}
\begin{equation}
\label{rani:4}
L\frac{dr}{dL} = 2(\epsilon' r + r^2).  
\end{equation}
There are two fixed points (i) $r=0$, and (ii) $r^*=-\epsilon'$.  A
nontrivial fixed point becomes important for $\epsilon'<0$, i.e., for
$d>1$.
See Appendix \ref{app:rani} for details.

We see certain similarities of  disorder or randomness becoming
marginally relevant at some dimension: $d_c=2$ for the random medium
problem, but $d_c=1$ for the RANI problem.  A new fixed point emerges
above this critical dimension.  For the random medium problem, this
implies the existence of a new phase and a disorder induced phase
transition, but for the random interaction problem it defines a new
type of critical behaviour.  These results based on the exact RG
analysis \cite{sm93:1,sm95:2} were later on also recovered from a
dynamic renormalization group study \cite{kalla}.

\subsection{Annealed case: three and four chains}
\label{sec:ann3}
With pairwise random interaction, the RANI problem can be formulated
for more than two chains also.  In such cases, even the annealed
averaging problem is not just the equivalent pure type
problem \cite{sm95:2}.  The Hamiltonian for a four chain system is
given by
\begin{equation}
{\sf H}=\frac{1}{2}\int_{0}^{N}dz\sum_{i=1}^{4}
\left (\frac{\partial{\bf{r}}_{i}(z)}{\partial z}\right
)^{^{\scriptstyle 2}} +
\int_{0}^{N} dz v_0(1+b(z))\sum_{ {i,j=1}\atop{i<j}}^4 \delta({\bf r}_{ij}(z)),
\end{equation}
where ${\bf r}_{ij}={\bf r}_i(z)-{\bf r}_j(z)$.  
After averaging of the partition function, using the Gaussian distribution of $b(z)$, one gets
the following effective Hamiltonian
\begin{eqnarray}
{\cal H}_{{\rm
eff}}&=&\frac{1}{2}\int_{0}^{N}dz\sum_{i=1}^{4}
\left (\frac{\partial{\bf{r}}_{i}(z)}{\partial z}\right
)^{^{\scriptstyle 2}} +
\bar{v}_0\int_{0}^{N} dz \sum_{{i,j=1}\atop {i<j}}^4
\delta({\bf r}_{ij}(z))
\nonumber \\ &&\quad -2v_0^2\Delta
\int_{0}^{N}\sum_{{i,j,k}\atop{i<j<k}}\delta({\bf r}_{ij})
\delta({\bf r}_{jk})
- 2v_0^2\Delta\int_{0}^{N} dz
\sum_{{i,j,k,l=1}\atop{i< j< k< l}}^4
\delta({\bf r}_{ij}) \delta({\bf r}_{kl}).\label{eq:h4}
\end{eqnarray}
The remarkable feature of the effective Hamiltonian is that
there are now two new types  of attractive interaction one
of which involves three chains,  while the other one couples
four chains together, similar to  the 2-chain  quenched problem.

If we take a three chain system, the corresponding effective
Hamiltonian will involve only the three chain term but no four chain
interaction of Eq. ~(\ref{eq:h4}).  There is now the possibility of a
disorder induced multicritical behavior.  The four chain attractive
interaction is marginal at $d=1$ and so is the three chain
interaction.  The presence of these two marginal operators (at $d=1$)
however remain decoupled mainly because for directed polymers, higher
order interactions (order $\equiv$ number of chains involved) do not
renormalize lower order interactions.  This has already been seen in the
overlap problem for the random medium case in Sec. \ref{sec:over}.
Therefore the resulting renormalization of the two new couplings are
independent of each other, and, in fact they are the same by virtue of the
nature of the interaction.  Because of the four body interaction, we
expect a disorder induced criticality as for the two chain quenched
case, but here this happens for a real four chain system - no replica
is involved.

\subsection{Unzipping}
\label{sec:unzip}
If we consider a pure problem, there will be an unzipping
transition
as described in
Sec. \ref{sec:unz_pure} above. The transition is first order in
nature.  In this particular case there is no thermal unbinding in one
dimension but if we forbid crossing of the polymers, then a thermal
unbinding is possible.  On a lattice this model can be solved exactly
with a phase boundary given by \cite{maren1}
\begin{equation}
  \label{eq:38}
 g_c(T)=T \cosh^{-1}\left ( \frac{1}{\sqrt{1-\exp(v/T)}-1+\exp(v/T)} 
-1\right ). 
\end{equation}
as shown in Fig. \ref{fig:dna}. One of the noteworthy features of the
phase diagram is the re-entrance in the low temperature
region \cite{maren1,maren2}.  At low temperatures, the unzipped chains
are stretched by the force so that on a lattice their entropy is
exponentially suppressed.  Though one gains energy by unzipping,  one
loses both the binding energy and the entropy of the bound chain. If
$\ln \mu_B$ is the entropy per site of the bound state in the
ground-state then, the transition occurs at $g_c= v +T \ln\mu_B$ with
a positive slope.  In the case of a single chain interacting with a
rod, the bound state has no entropy ($\mu_B=1$) and therefore no
re-entrance.  This argument also shows that the transition is first
order.

\figuz

It has been argued that the unzipping
transition
for the quenched averaged RANI
case is second order \cite{lub}.  However for real DNA, it is not the
quenched averages that matter. There is strong ensemble dependence and
sample to sample variation.  This has been exploited to identify point
mutants by a comparison of the unzipping force in a fixed distance
ensemble \cite{muta}.
An experimental
determination of the unzipping phase boundary for a real  DNA has been
reported in Ref.  \cite{danil}

\section{Hierarchical lattice}
\label{sec:hier}
Useful information can be obtained from models amenable to exact
analysis even if they look artificial.  Real space renormalization
group 
approach can be handled
in an exact fashion for a class of tailor-made lattices called
hierarchical lattice.  Such lattices are constructed in a recursive
fashion as shown in Fig. \ref{fig:1}.  The problem of a directed
polymer in a random medium on hierarchical lattices has been
considered in Ref.  \cite{cook,med}.  Here we consider the RANI problem
on hierarchical lattices. As already noted, the effective
dimensionality is $d_{\rm eff}=(\ln 2b)/\ln 2$ for a motif of $2b$
bonds.  Two different situations can arise, the polymers interacting
on shared bonds \cite{sm:har} or on contacts at vertices \cite{smb:div}.
Two cases are different.  Aperiodic variation of the interaction has
also been considered \cite{hadad}.

\subsection{Randomness on sites}

Let us now consider interaction on the vertices.  The problem is
different from the bond case because by construction there are sites
with large degrees.  In other words, all sites are not equivalent,
unlike the bond case where all bonds are on the same footing.

Let $Z_{\mu}$ be the partition function of a given realization of
randomness and let $S_{\mu}=b^{L_{\mu}-1}$ the number of single-chain
configurations, at the $\mu$th generation. Here $L_{\mu}=2^{\mu}$ is
the length of a directed polymer. We define ${\cal
Z}_{\mu}(n)=\dav{Z^{\nu}_{\mu}}/S_{\mu}^{2n},$ to factor out the free
chain entropy. We call ${\cal Z}_{\mu}(n)$ the moments.  For a given
temperature, there is a critical value $n=n_c(\overline y)$ below
which all moments are in their high temperature phase, in the
thermodynamic limit of course (overbar indicating disorder average).
In this limit, ${\cal Z}_{\mu}(n)$ approaches a fixed point value for
$n < n_c$, whereas, for $n >n_c$, the moments diverge but with a
finite ``free energy" density $f_{\mu}(n)=(nL_{\mu})^{-1} \ln {\cal
Z}_{\mu}$.  The approach to the thermodynamic limit can be written
generically as
\begin{equation}
  \label{eq:34}
{\cal Z}_{\mu}(n)^{1/n}={\cal Z}(n)^{1/n}+B_Z(n)L_{\mu}^{-\psi}+ ...  
\end{equation}
where ${\cal Z}(n)$ is the thermodynamic limit $(\mu\rightarrow
\infty)$ and $B_Z(n)$ is the amplitude of the finite size correction.

For a given realization of disorder, the partition function can be
written as (see Fig.1)
\begin{equation}
  \label{eq:35}
Z_{\mu+1} =bZ_{\mu}^{(A)}yZ_{\mu}^{(B)}+b(b-1)S^4_{\mu}  
\end{equation}
The first term originates from the configurations that require the two
directed polymers to meet at a vertex.  In contrast, the second term
counts the ``no encounter" cases. There are no energy costs at the two
end points.  The Boltzmann weight is random and for a Gaussian
distribution of energy, $\overline{y^m}={\overline{y}}^{m^2}$. The
moments of the partition function can be written as
\begin{equation}
  \label{eq:36}
  {\cal Z}_{\mu+1} (n) =b^{-n}\sum^n_{m=0} P_{nm}Z^2_{\mu}(m)\qquad
  {\rm with} \ P_{nm}={n\choose m}(b-1)^{n-m}{\overline{y^m}},
\end{equation}
and the initial condition $Z_0(n)=1$ for all the moments because there
is no interaction in the zeroth generation (one single bond).

By iterating the recursion relations, the moments can be calculated
exactly to arbitrary precision from which $B_Z(n)$ can be estimated.
One finds that $B_Z(n)$ blows up as
\begin{equation}
  \label{eq:37}
B_Z(n) \sim (n_c-n)^{-r}   \ {\rm as}\ n\rightarrow n_c-,
\end{equation}
with $r=0.71\pm 0.02$.  This exponent is independent of temperature
but depends on $b$.

As mentioned, in the site version there are special sites with large
degree.  Whether such vertices lead to a Griffiths' type singularity
needs to explored. Griffiths singularity in 
the context of random interactions has recently been discussed in
Ref.  \cite{kafri}.

\subsection{Bond randomness: Problem with Harris criterion}
According to the Harris criterion, disorder is irrelevant at a
critical point if the pure specific heat
exponent
$\alpha<0$.  We show that a more general criterion is required for the
directed polymer problem \cite{sm:har}.

We place two interacting polymers on a hierarchical lattice.  They
start from one end (bottom) and meet at the other end (top). See Fig.
\ref{fig:1}.  There is an attractive interaction $-v (v>0)$ if a bond
of the lattice is shared by two polymers.  As in real space, here also the
polymers undergo a binding unbinding transition for $b >2$.

Randomness is introduced by allowing the interaction energy to be
random on each and every bond. The first model, model A, has
independent random energy on all the $2b$ bonds. The randomness in the
second model, model B, is taken only along the longitudinal direction
so that equivalent bonds on all directed paths have identical random
energy.  Model B is a hierarchical lattice version of the continuum
RANI model.

The pure problem can be solved easily by a Real space renormalization
approach 
where one needs only
the renormalization of the Boltzmann factor $y=\exp ({v/ T})$.  Let
$y_n, Z_n$, and $E_n$ be the renormalized weight, partition function
and energy at the $n$th generation. By decimating the diamonds the
recursion relations are given by
\begin{eqnarray}
  \label{eq:26}
 y_{n-1}&=&(y^2_n+b-1)/b,\\ 
Z_{n+1}&=&(Z^2_n+b-1)/b,\\
E_{n+1} &= &\frac{2}{b}\ \frac{Z^2_n E_n}{ Z_{n+1}}. \label{eq:en}
\end{eqnarray}
The two fixed points of the quadratic recursion relation,
Eq. (\ref{eq:26}), are $1$ and $b -1$ of which the larger one is the
unstable fixed point representing the transition point. Since $y > 1$
there is a transition at $y_c=b-1$ only for $b > 2$. The other fixed
point $y^* =1$ corresponds to the high temperature limit.

The length scale exponent
$\zeta$, and the specific heat
exponent, $\alpha$, can be obtained from Eqs. (\ref{eq:26})-(\ref{eq:en})
as
\begin{equation}
  \label{eq:27}
 \zeta = \frac{\ln \ 2}{\ln [2(b-1)/b]} , \ {\rm and} \ \alpha = 2-\zeta.
\end{equation}
Note that $\alpha < 0$ for $b <2+\sqrt 2$.  It is clear that
hyperscaling holds good with $d = 1$ and not the effective dimension
$d_{\rm eff}$ of the lattice. It is also gratifying to observe that
the same $\nu$ and $\alpha$ describe the finite size scaling form of
$E_n$.

For the disordered case, the recursion relation for the Boltzmann
weight can be written as
\begin{equation}
  \label{eq:28}
  y_{n-1}=b^{-2}(y^{(11)}y_n^{(12)}+y^{(21)}y_n^{(22)}+...
y^{(b1)}y_n^{(b2)})+(b-1)b^{-1},
\end{equation}
where $y^{(ij)}_n$ is the Boltzmann weight in the nth generation for
the disorder on the upper $(j=1)$ or lower $(j=2)$ part of the $i$th
branch as in Fig. \ref{fig:1}. To understand the effect of the
disorder at the pure critical point we introduce a small disorder
$y^{(ij)}_n=y_c+\varepsilon^{(ij)}_n$. The average of the disorder,
$\dav{\varepsilon}$, acts like the temperature as it measures the
deviation from the pure critical point.  The second moment is the
measure of disorder. In principle, one should look at the variance of
$\varepsilon$ , but at the pure critical point the variance would be
the same as the second moment. Since our motivation is to find the
flow of the disorder at the pure fixed point, we need only study the
first two moments, starting from a finite and small variance.

The crossover exponent
for the disorder is
defined through the homogeneity of the singular part of the free
energy in terms of the scaling fields $\mu_1$ (temperature) and
$\mu_2$ (disorder). Under decimation, the free energy behaves
\begin{equation}
  \label{eq:29}
f_{\rm sing}(\mu_1,\mu_2)=\mu_1^{2-\alpha}\ {\sf f}(\mu_2/\mu_1^{\phi})
\end{equation}
defining $\phi$ which can be obtained from the RG recursion relations
for the first two moments.  This crossover exponent determines the
relevance of disorder at the critical point and can be computed
exactly for both models A and B.  One finds a striking difference
between the two models as
\begin{equation}
  \label{eq:30}
 \phi =\frac{\ln(2y^2_c/b^3)}{\ln[2(b-1)/b]} \quad ({\rm model\  A}),
 \quad {\rm and}\ 
\phi= \frac{\ln[2(b-1)^2/b^2]}{\ln [2(b-1)/b]} \ {\rm (model \  B)}.
\end{equation}
For model A, $\phi$ is negative for all $b >2$, implying irrelevance
of disorder and $\phi\neq \alpha$ but it is equal to $2-d_{\rm
eff}\nu$, while for model B, $\phi=\alpha$ and not $2-d_{\rm eff}\nu$.
Since the randomness in model B is highly correlated, the Harris
criterion is less expected to be valid here as opposed to model A.
but  it turns out to be so.

In order to construct a general framework for checking the validity of
the Harris criterion,
we start with the Taylor
expansion of the recursion relation of Eq. (\ref{eq:28}),
\begin{equation}
  \label{eq:31}
\varepsilon
=g_s(b)(\varepsilon_1+\varepsilon_2+...\varepsilon_{\cal
  N})+0(\varepsilon^2_i)+...,    \quad({\cal N}=2b) 
\end{equation}
which defines $g_s(b)$. Simple arguments show that $g_s(b)$ determines
$\alpha$ whose positivity is guaranteed if $g_s(b) >\sqrt 2{\cal
N}^{-1}$ Now, suppose that the bonds are grouped in $n$ sets with
$N_i$ bonds in the ith group such that the members of a set have the
same randomness. Obviously $\sum N_i={\cal N}$.  Starting with a
narrow distribution, the relevance of the disorder at the pure
transition then requires
\begin{equation}
  \label{eq:32}
  g_s(b) >(N^2_1+N^2_2+....+N^2_n)^{-1/2}.
\end{equation}
Hence, the Harris criterion holds good if either
\begin{equation}
  \label{eq:33}
g_s (b) > \max \left[\frac{\sqrt 2}{\cal N}, \frac{1}{\sqrt{\sum N^2_i}}\right],
\ {\rm or}\ g_s (b) < \min \left[\frac{\sqrt 2}{\cal N}, \frac{1}{\sqrt{\sum N^2_i}}\right].
\end{equation}
 For model A, ${\cal N}=2b$ and $N_i=1$ $\forall i$, while for model
B, $n = 2$ with $N_i=b$.

If the disordered models are classified by $\pm$ according to the sign
of $\alpha$, and I or R according to irrelevance or relevance of disorder, then
the Harris criterion
suggests the existence of
only two classes (+R) and (-I). On the other hand, the above
inequalities allow special classes like (+I) and (-R) where the Harris
criterion fails. Model A is in the (+I) class for $b>2+\sqrt 2$. Model
B is in either the (+R) or (-I) class depending on $b$.  It is
possible to construct models that would belong to any of the four
possible classes, especially (-R) \cite{sm:har}.

\section{Summary}
\label{sec:summ-open-probl}
The behaviour of a directed polymer in a random medium in 1+1
dimensions seems to be well understood.  There is a strong disordered
phase at all temperatures for $d<2$.  For $d>2$ renormalization group
analysis shows a phase transition from a low temperature strong
disordered phase to a weak disorder, pure-like phase.  There are rare
configurations with degenerate widely separated ground states, giving
a contribution to ``overlap'', and strong sample dependent
response to an unzipping force.

The RANI model remains less understood compared to the random medium
problem. Exact renormalization analysis establish the marginal
relevance of the disorder at $d=1$, indicating a disorder dominated
unbinding transition in $d\ge 1$.   Several features including  a 
generalization of the Harris criterion for this  criticality via relevant disorder
and aspects of unzipping have been discussed.  

In both cases of random medium and random interaction,  many issues still remain open.

\noindent {\bf{Acknowledgments}}

The author thanks Rajeev Kapri and Soumen Roy for many useful comments
on the manuscript. 

\vspace{1cm}

\noindent {\bf{APPENDIXES}}

\appendix

\section{Typical vs. average}
\label{app:typ}
Consider a random variable $x$ that takes two values
\begin{subequations}
\begin{eqnarray}
\label{eq:rar1}
X_1&=& e^{\alpha \sqrt{N}} \ {\rm and} \ X_2= e^{\beta N}, \beta >1,\\
\lefteqn{\mathrm{with\   probabilities}\hspace{2cm}}\nonumber\\
p_1&=&1-e^{-N}, \ {\rm and} \ p_2=e^{-N}.
\label{eq:rar2}
\end{eqnarray}
\end{subequations}
In the limit $N\rightarrow \infty$, the average value $\dav{x}
\rightarrow e^{(\beta-1)N}$  while the typical or most probable value
is $x=X_1$ with probability $1$.   On the other hand $\dav{\ln x}
\rightarrow \alpha \sqrt{N}$ in the same limit, showing that $\dav{\ln x}$
is determined by the typical value of the variable while the moments
are controlled by the rare events.  Note that this peculiarity
disappears if $x$ has a smooth probability distribution in the sense
of no special or rare events.

\vspace{2\parskip}

\section{Pure Polymers}
\label{app:pure}
The universality of the ``Gaussian'' polymer is actually a consequence
of the Central limit theorem.  Suppose we construct a flexible polymer
from bonds with independent probability distribution $\psi({\bf r})$
for a bond vector ${\bf r}$.  The end-to-end distance is given by
${\bf R}=\sum_i{\bf r}_i$ so that the probability density of ${\bf R}$
is
\begin{eqnarray}
\label{eq:appa.1} 
P({\bf R})&=&\int \prod_i d{\bf r}_i \ \psi(r_i) \delta (\sum_i{\bf
  r}_i - {\bf R}) \nonumber\\ 
&=&\int \prod_i d{\bf r}_i \ \psi(r_i) \exp(i{\bf k}\cdot{\bf r}_i)
\exp(-i{\bf k}\cdot{\bf R}) d{\bf k}\\ 
&=& \int [{\hat{\psi}}({\bf k})]^N \exp(-i{\bf k}\cdot{\bf R}) d{\bf k},
\end{eqnarray}
where $\hat{\psi}({\bf k})$ is the Fourier transform of
$\psi({\bf_r})$.  For a symmetric distribution with finite variance,
$\ln \hat{\psi}({\bf k}) \approx 1-A k^2/2 ...$, which on integration
over $k$ leads to a Gaussian distribution.  This is valid for a lattice
model also. 

With the probability distribution of Eq. (\ref{eq:1.2}), the entropy in
a fixed distance ensemble can be written as
\begin{equation}
\label{eq:appa.2}
S({\bf r}) = S(0) - \frac{1}{2} \ \frac{r^2}{R_0^2},
\ {\rm which\  gives} \ F({\bf r})= F(0) + \frac{1}{2} \ \frac{T r^2}{R_0^2}.
\end{equation}
This identifies an effective spring constant for the polymer, namely
$3T/2R_0^2$.  This spring like behaviour is purely an entropic effect.

An important and general point is to be noted.  The macroscopic
quantity  involves an ``external'' parameter like ${\bf r}$ which is
scaled by $R_0$ the characteristic long-length scale size of the
polymer.  That the long distance behaviour can be described by a
single length scale is the basis of ``scaling'' approach to polymers.

Another approach to scaling is to study the changes in the properties
of a polymer as the microscopic variables are scaled.  E.g., if we
make a scale change, ${\bf r} \rightarrow b {\bf r}$ and
$z\rightarrow b^{1/\nu} z$, the Hamiltonian of Eq. (\ref{eq:1.1}) remains
invariant if $\nu=1/2$.  Under such a transformation, the size behaves
as
\begin{equation}
  \label{eq:58}
  R_0(N) = b^{-1}R_0(b^{1/\nu} N),
\end{equation}
so that by choosing $b=N^{-\nu}$, $R_0\sim N^{\nu}$, i.e., the size
exponent becomes $\nu=1/2$.  In presence of interaction or disorder,
$\nu$ may not be obtained so simply and then renormalization group
methods become useful.

\subsection{Scaling approach in presence of a force}
If the polymer is now pulled with a force $g$, keeping the end at
$z=0$ fixed, the polymer would align on the average with the
force. The polymer can be thought of as consisting of blobs within
each of which a polymer can be considered as free from the force
whereas the blobs as unit form a still coarse grained model that shows
stretching. This is shown schematically in Fig. \ref{fig:blob} and is
used in Sec. \ref{sec:response-force}.  This  description is called a
``blob picture''.  This picture essentially depends on the scaling
idea that $R_0$ is the only relevant scale for the macroscopic
description.  This can definitely be justified at the Gaussian level.
The partition function with the unzipping force can be written as
\begin{equation}
Z=\int d{\bf r} P({\bf r},N) \exp(\beta {\bf g}\cdot{\bf r}),
\end{equation}
where $P({\bf r},N)$ is Eq. (\ref{eq:1.2}).  The Gaussian integral can
be done (keeping $T$ explicitly in Eq. (\ref{eq:1.2})) to obtain
\begin{equation}
\label{eq:appa.3}
|\!<{\bf r}>\!|= \frac{R_0^2}{T}\, g,
\end{equation}
which is consistent with the idea of an effective spring constant of
the polymer derived after Eq. (\ref{eq:appa.2}).

A scaling approach to derive Eq. (\ref{eq:appa.3}) would as follows.
Let $R_g$ be the characteristic size of the polymer in presence of the
force. Then, from a dimensional analysis point of view, this is
similar to the zero force size $R_0$. From the nature of the force
term, $g$ is dimensionally like $T$ divided by a length scale.  Only
lengthscale in the problem is $R_0$.  Hence a dimensionally correct
form is
\begin{equation}
R_g\sim R_0 \ {\sf f}(gR_0/T).
\end{equation} 
Note the absence of any microscopic scale in the above form.  All
microscopics go in $R_0$. For a linear law at small force we require
${\sf f}(x) \sim x$ ($x\rightarrow 0$) giving back
Eq. (\ref{eq:appa.3}). One may rephrase this by saying that the force
has a characteristic size $\xi_g \sim T/g$.

If the polymer is confined in a tube of diameter $D$ then the
dimensionless variable is $R_0/D$.  This will appear in the form for
change in entropy or in ``confinement energy''.  This is used in
Sec. \ref{sec:confine}.

\vspace{2\parskip}

\section{Self-averaging}
\label{app:sa}
Let us build a large system by adding blocks A, B, C, D, ...
systematically so that an extensive quantity is a sum over its values
on individual blocks. In case this averaging over blocks leads to a
very sharp probability distribution, then no further disorder averaging
is warranted, i.e., any large sample would show the average behaviour.
A quantity with this property is often called {\it
self-averaging}.
This may not be the case if
the distribution is broad especially in the sense discussed in App. A.
A self-averaging quantity has the advantage that one may study one
realization of a large enough system without any need of further
disorder or sample averaging. For numerical simulations, the
statistics of a non-self-averaging quantity cannot be improved by
increasing the number of realizations.


To be quantitative, let us choose a quantity $M$ which is extensive
meaning $M=N m$ where $m$ is the ``density'' or per particle value.
This is based on the additive property over subsystems $M=\sum M_i$.
For a random system we better write $M=M(N,\{Q\})$, with $\{Q\}$
representing all the random variables.  To recover thermodynamics, we
want $\dav{M}$ to be proportional to $N$ for $N\rightarrow\infty$.
Now, if it so happens that for large $N$
\begin{equation}
\label{eq:2.2.1}
M(N,\{Q\}) \rightarrow N m_d,
\end{equation}
with $m_d$ independent of the explicit random variables, then $M$ is
said to have the self averaging property.  Note that no averaging has
been done in Eq. (\ref{eq:2.2.1}).  One way to guarantee this
self-averaging is to have a probability distribution
\begin{equation}
  \label{eq:23}
  P(M/N) \stackrel{N\rightarrow\infty}{\longrightarrow} 
\delta(m_d).
\end{equation}
This is equivalent to the statement that the sum over a large number
of subsystems gives the average value, something that would be
expected in case the central limit theorem (CLT) is applicable.  This
generally is the case if quantities like $M$ for the sub-blocks are
independent and uncorrelated random variables.

For many critical systems CLT may not be applicable and self-averaging
is not self-evident.  A practical procedure for testing self-averaging
behaviour of a quantity $X$ is to study the fluctuations $\sigma_N^2
\equiv \dav{X^2} - {\dav[2]{X}}$ and then check if
\begin{equation}
\label{eq:2.2.2}
R_{X,N}=\frac{\sigma_N^2}{\dav[2]{X}} \rightarrow 0, \ {\rm as\ }
N\rightarrow \infty.
\end{equation}
A quantity is not self-averaging if the corresponding $R$ does not
decay to zero.  The central limit theorem would suggest $R_{X,N}\sim
N^{-1}$, while a decay of   $R_X$ slower than this would be termed as
```weakly'' self-averaging. 

 We may then classify a quantity $X$, based on the large $N$
 behaviour,  as follows:

\vspace{\baselineskip}
\begin{tabular}{lll@{\vspace{\baselineskip}}}
\label{eq:strng}
$R_{X,N}$ & $\sim$ constant                & $\Rightarrow$  {\it non} self-averaging\\
         & $\sim\  \displaystyle{ N^{-1}}$ & $\Rightarrow$ {\it strongly}  self-averaging\\
         & $\sim N^{-p}\ {\rm with} \ p<1 $ & $\Rightarrow$ {\it weakly} self-averaging.
\end{tabular}

\vspace{\baselineskip}

Recent renormalization group arguments seem to suggest that if
disorder is relevant then at the new (disorder-dominated) critical
point thermodynamic quantities are not self-averaging \cite{AH-PRL-96}.
The arguments leading to this extremely significant prediction of non
self-averaging nature of critical quantities can be summarized as
follows.  

Let $\tilde{t_i}=|T-T_c(i,N)|/T_c $ be a sample dependent reduced
temperature where $T_c(i ,N)$ is a pseudo-critical temperature of
sample $i$ of $N$ sites with $T_c$ as the ensemble averaged critical
temperature in the $N\rightarrow\infty$ limit.  In terms of this
temperature, a critical quantity $X$ is expected to show a sample
dependent finite size scaling form
\begin{equation}
\label{fss}
X_i(T ,N)=N^\rho \, Q (\tilde{t_i}N^{1/{d\,\zeta}} )
\end{equation} 
where $\rho$ characterizes the behaviour of $\dav{X}$ at $T_c$,
 $\zeta$ being the length scale exponent.
(E.g. $\rho=\gamma/{d\zeta}$ when $X$ is the magnetic susceptibility
$\chi$.  This is plausible because the critical region sets in when the
size of the system is comparable to the correlation volume 
$\xi^d$ which diverges  as $|T-T_c|^{-\zeta}$.  The RG approach
validates this hypothesis of Eq. (\ref{fss}), especially the
absence of any extra anomalous dimension in powers of $N$ for $R_X$.
Incidentally, this hypothesis, Eq. (\ref{fss}), excludes rare events
that may lead to Griffiths' singularity.  Using this scaling form, the
relative variance $R_X$ at the critical point or in the critical
region can be determined as
\begin{equation}
\label{eq:sa-2}
R_X \sim [(\delta T_c)^2] \, N^{2/d{\zeta}} ,
\end{equation}
where $[(\delta T_c)^2]$ is the sample average variance of the
pseudo-critical temperature.  A finite size scaling form for $R_X$
can also be written down, but it is not required here.

A random system can have several temperature scales, namely
$(T_c(N)-T_c)$ and $(T-T_c)\sim \xi^{-1/\zeta}$, in addition to the
shift in the transition temperature itself.  For a system with {\it
relevant disorder},  all these scales should behave in the
same way so that typical fluctuations in the pseudo-critical
temperature is set by the correlation volume ($\xi^d$).  In the finite
size scaling limit $\xi^d \sim N$, and then 
\begin{equation}
\label{eq:sa-3}
[(\delta T_c)^2] \sim N^{-2/{d{\zeta}}} \qquad {\rm (relevant \ disorder)}
\end{equation} 
An immediate consequence of this is that $R_X$
approaches a constant as $N\rightarrow\infty$ indicating {\it complete
absence of self-averaging} at the critical point in a random system.

For a pure type critical point ({\it irrelevant disorder}) $\alpha <0$
where $\alpha$ is the specific heat exponent (i.e. $c\sim
|T-T_c|^{-\alpha}$) of the pure system.  In this case the fluctuation
in $T_c$ is set by the size, i.e.
\begin{equation}
\label{eq:sa-4}
[(\delta T_c)^2] \sim N^{-1} \qquad {(\rm irrelevant \ disorder)}
\end{equation} 
so that, by using the hyperscaling relation $2-\alpha=d\zeta$, one
gets $0<p=|\alpha/{\bar{\zeta}}|<1$ where $R_x\sim N^{-p}$.  Hence all
critical quantities in this case are weakly self-averaging.  Moreover,
it is the {\it same power law} involving $\alpha$ and $\zeta$, for every
critical quantity $X$ no matter what its critical exponent is.

These predictions have been verified for various types
relevant and irrelevant disorders \cite{WD-PRE-98,Dillmann,Marques1}.
Exception to such non self-averaging behaviour with relevant disorder
occurs if the $T_c$ distribution approaches a delta function for large
lattices. In such a situation, one gets back strong self-averaging
behaviour \cite{rb04}.

\section{Details of RG for polymers: dimensional regularization}
\label{app:rg}
Some details of the renormalization group approach for polymers as
done in Sec. \ref{sec:zrg} are given here \cite{jj}. We consider the
problem of two interacting directed polymers and study the second
virial coefficient.  The second virial coefficient is related to the
two-chain partition function with all the ends free.  Dimensional
regularization is to be used here.

For this appendix we take a simpler form of the Hamiltonian given be
Eq. (\ref{eq:5}) as
\begin{equation}
  \label{eq:39}
   H_2= \frac{d}{2} \int_0^N dz \sum_i^2\left( \frac{\partial {\bf
        r}_i}{\partial z}\right ) ^{^2} 
        - v_0  \int_0^N d z \  \delta({\bf r}_1(z)-{\bf r}_2(z)),
\end{equation}
where $v_0$ is the bare or starting interaction strength.  By
introducing an arbitrary length scale $L$ (may be the scale chosen to
study the system or in momentum shell approach, this is the cutoff
length), we may define the dimensionless variables $u_0=v_0
L^{\epsilon}/(2\pi)^{d/2}$ and ${\sf N}=N/L^2$, with $\epsilon=2-d$.
For long distance properties, we want $L$ to be large.

By definition, the second virial coefficient comes from the connected
two chain partition function with all the ends free.  An expansion in
terms of the coupling constant would involve polymer configurations as
shown in Fig. \ref{fig:ufp}.  This is incidentally identical to Eq.
(\ref{eq:2.1.3a}).  Each line represents  the probability of free
polymer going from ${\bf r},z$ to ${\bf r}',z$,
\begin{equation}
  \label{eq:40}
  G({\bf r}',z'|{\bf r}, z) = \frac{1}{[2\pi (z'-z)]^{d/2}}
 \  \exp\left(- \frac{({\bf r}'-{\bf r})^2}{2(z'-z)}\right)
\end{equation}
as in Eq. (\ref{eq:60}).  A crossing of the lines in a diagram  represents an
interaction at $({\bf r},z)$ which can take place anywhere requiring
an integration over ${\bf r}$ and $z \in (0,N)$.

\figolp

Since $G$ is normalized, the spatial  integration over the free end
points lead to unity and so the dangling legs of the diagrams do not
require any evaluation.  
One needs to do only the loop integrals.  For example,  the one loop
diagram shown in Fig. \ref{fig:olp} corresponds to 
\begin{eqnarray}
  \label{eq:68}
&&  \int_0^N dz'\ \int_0^{z'} dz \int d{\bf r}\int d{\bf r'} 
    \ G^2({\bf r}'z'|{\bf r}z) \nonumber\\
&&\qquad \times    \int d{\bf r}_1 \ G({\bf  r}z|{\bf r}_10)
      \int d{\bf r}_2  \ G({\bf r}z|{\bf r}_20)
      \int d{\bar{\bf r}}_1 \ G({\bar{\bf r}}_1N|{\bf r}'z')
     \int d{\bar{\bf r}}_2\ G({\bar{\bf r}}_2N|{\bf r}'z'),
\end{eqnarray}
which after integrations over the end coordinates ${\bf r}_1, {\bf
  r}_2, {\bar{\bf r}}_1, {\bar{\bf r}}_2$ reduces to  an  integral of the type 
\begin{equation}
  \label{eq:64}
  \int_0^N dz'\ \int_0^{z'} dz \int d{\bf r}\int d{\bf r'} G^2({\bf
    r}'z'|{\bf r}z)
\end{equation}
so that integrations over the space coordinates lead to integrals of
the type 
given by Eqs.\ref{eq:1.5a}, \ref{eq:2.1.4}.
Using $d$ as a continuous variable, one can write 
\begin{equation}
  \label{eq:65}
 \int_0^N\ dz \ z^{-\Psi}= \frac{N^{1-\Psi}}{1-\Psi} 
\end{equation}
with $\Psi=d/2$.  This form is now valid for all
$d$ so that the loop integral may be written as, besides other
constant factors, 
\begin{equation}
  \label{eq:66}
  2N {\cal V}\Delta^2\ \frac{N^{1-d/2}}{2-d}
\end{equation}
with a simple pole at $d=2$. ${\cal V}$ is the total volume.  One then identifies $d=2$ as the
special dimension. 

Using the above form with the singularity at $d=2$, we may write the
second virial coefficient as
\begin{equation}
\label{eq:43}
A_2 = 2\pi {\sf N} {\cal V} u_0 L^{2-\epsilon} \left [ 1 + u_0 (1 +({\epsilon}/2) \ln
{\sf N}+ ...) 
 \frac{2}{\epsilon}  + ...  \right ], 
\end{equation}
where $a$ is a constant, and we used 
${\sf N}^{\epsilon/2} = 1 +({\epsilon}/2) \ln {\sf N}+ ...$.  i
The series has problem at
$d=2$.  We try to absorb the divergences by redefining all the
parameters in hand, in this particular case, only $u_0$ and ${\sf N}$.
It is clear the the source of divergence is the region when $z,z'$ are
close by.  On a bigger scale it is these close by reunions which would
contribute to the effective interaction seen.  It is therefore natural
that these divergences ultimately determine the RG flow of the
coupling constant.

Define a {\it renormalized} coupling constant as 
\begin{equation}
  \label{eq:67}
u_0 = u + D_1 u^2 + D_2 u^2 + .... = u Z_u.  
\end{equation}
Substitute in the expression for $A_2$, and
choose $D$'s to cancel the poles. $Z_u$ is called a {\it
multiplicative renormalization factor}.

We adopt the minimal subtraction scheme
where the $D$'s are chosen to subtract
the poles and only the poles.  The calculation is order by order and
so, to one loop order, one cannot determine $D_2$, which involves
$u^3$ (two loop term ).

Choosing $D_1 = - 1/\epsilon$, we see that the divergence is
removed to  O$(1/\epsilon)$.  Upto this order, it follows that 
\begin{equation}
\label{eq:41}
 u_0 = u - \frac{u^2}{\epsilon} + ... 
\end{equation}
Since this absorbs all the divergences, we do not have to renormalize
${\sf N}$. The divergence-free quantity is $A_{2R}(L,u,N)$ (R for
renormalized; expressed in terms of $u$).  Now, $A_{2}$ should not
depend on $L$ because $L$ is put in by hand. This is ensured by
demanding that
\begin{equation}
\label{eq:42}
L\frac{d A_{2}}{dL} = 0,
\end{equation}
where the factor of $L$ in front has been put in for a dimensionless
derivative operator. Written in a long form
\begin{equation}
\label{eq:44}
\left [L\frac{\partial}{\partial L}
+ \ \beta(u) \frac{\partial}{\partial u} \right ]A_{2R}(L,u, N) =
0,\quad \quad {\rm with} \quad \beta(u) = L \frac{\partial u}{\partial L}.
\end{equation}
Note that  $A_{2,R}(u) = A_2(u_0)$, though, in general, additional
renormalization factors (multiplicative and/or additive) may be needed.

By definition, $\beta(u)$ tells us how the renormalized $u$ changes
with the length scale and is called the beta function in RG.  Some
more algebra gives
\begin{equation}
\label{eq:45}
\beta(u) \equiv L\frac{\partial u}{\partial L} =\frac{\partial u}{\partial
u_0}\ L  \frac{\partial u_0}{\partial L} = \epsilon u_0 \frac{\partial
u}{\partial u_0} = \epsilon \left(\frac{\partial \ln u_0}{\partial
u}\right)^{^{\scriptstyle -1}} 
\end{equation}
so that the variation of the coupling constant with scale $L$ is given
by 
\begin{equation}
  \label{eq:77}
 L\frac{\partial u}{\partial L}= \beta(u)\equiv  \epsilon \, u + u^2/(2\pi).
\end{equation}
The factor of $2\pi$ can be absorbed in the definition of $u$, as have
often done.   This equation, Eq. \ref{eq:77} is called a
renormalization group flow equation 
with the initial condition $u=u_0$ for some $L=L_0$.  It
is analytic in $\epsilon$ so that various dimensions can be handled
with this equation.  Initial condition may be taken as $u=u_0$ for
some $L=L_0$.  In this particular case the beta function is exact to
all orders.

The zeros of the beta function are called fixed
points
which can be of two
types, stable or unstable.  For this particular case, the fixed points
are $u^*=0$ and $u^*=- \epsilon$.   If we start with a very small $u$,
the flow equation $L du/dL \approx \epsilon u$ shows a growth of $u$
with $L$ if $\epsilon >0$, i.e., if $d<2$.  This means $u$ is a
relevant variable at the noninteracting point.  For the disordered
system it translates to relevance of disorder at the pure fixed point.

One gets a nontrivial fixed point at $u^*=- 2\pi\epsilon$ where
$\beta(u) =0$ which is an unstable fixed point if $\epsilon <0$.  This
unstable fixed point for $d>2$ represents the disorder induced
critical point.

For the disorder problem, $u<0$ is not meaningful, but for the
interaction problem as in the RANI model, full range of $u$ is
allowed.  The nontrivial fixed point for $d<2$ is a stable fixed point
and it describes the phase of repulsive polymers (fermion like).
There is a critical binding-unbinding transition for $d>2$ for pure
attractive short range interaction.  The unbinding transition is at
zero potential for $d\le 2$

For the transition one can define a characteristic length $\xi$ so
that at scales$>\xi$, the critical effects can be ignored.  If we
start with an initial value $u_0=u^*+\Delta_0$, then one may say this
crossover happens at some arbitrarily chosen value of $\Delta u = 1$
say where $\Delta u$ is the renormalized deviation from the fixed
point.  One may determine this by linearization as
\begin{equation}
\label{eq:46}
L\frac{d\Delta u}{dL} = \mid \epsilon\mid \Delta u,
\end{equation}
which gives $\Delta u = \Delta_0 (L/L_0)^{\mid\epsilon\mid}$.  If we
take $\Delta_0<<1$ as a measure of the deviation from the critical
point (like $T-T_c$), then a small deviation grows and goes over
either to the stable f.p. for the unbound phase or to $-\infty$ for
the bound phase depending on the starting initial sign.

Setting $L=\xi$, we then get 
\begin{equation}
\xi = \mid \Delta_0\mid^{-\zeta}, \ {\rm and}\ 
\zeta=\frac{1}{\mid\epsilon\mid}.
\end{equation}
Since the beta function is exact,  we have obtained the exact
correlation length scale exponent for the binding-unbinding critical
transition.

At the critical dimension, $d=2$, the flow equation is 
 \begin{equation}
   \label{eq:47}
   L\frac{d\Delta u}{dL} =  (\Delta u)^2,
 \end{equation}
which gives $\xi\sim \exp(1/\Delta_0)$. This exponential dependence of
$\xi$ on $\Delta_0$ accounts for the divergence of $\nu$ at
$\epsilon=0$.

\section{RG of the KPZ equation: momentum shell technique}
\label{app:kpz}
We discuss the momentum-shell RG approach to the KPZ equation
 \begin{equation}
\label{eq:kpz1}
\frac{\partial F}{\partial z} = \frac{T}{2K} \nabla^2 F - \frac{1}{2K}
(\nabla F)^2 + \eta({\bf r},z). 
\end{equation}
which is Eq. (\ref{eq:kpz}).
 More details on this equation  may be found in Ref.  \cite{barab}.  The
idea is to get the behaviour of the parameters of the differential
equation in the long distance limit.  The three parameters are $T$,
$K$ and $\Delta$, of which $K$ remains invariant so that
$2\nu=\theta+1$, Eq.  (\ref{eq:15}), is satisfied.  Note that if the
nonlinear term (i.e. the force term in the fixed distance ensemble) is
absent, the differential equation becomes linear which can be solved
exactly\cite{ew}.  The RG scheme uses this exact solution for an iterative
approach to tackle the nonlinear term.  In contrast to the RG in the
polymer approach of App. \ref{app:rg}, here the starting point is not
a Gaussian polymer but a linear equation which need not represent any
polymer.

It is convenient to work in the Fourier space
\begin{equation}
  \label{eq:49}
  F({\bf x},z)=\int_{-\infty}^{\infty}
  \frac{d\omega}{2\pi}\int_{k<\Lambda} \frac{d^d{\bf k}}{(2\pi)^d}
  F({\bf k},\omega)\equiv \int_{\bf q} \int_{\omega}\ F({\bf k},\omega)
\end{equation}
where $\Lambda$ is  an upper cutoff in $k$-space (related to  a real
space  short distance  cutoff).  The long chain, long distance limit
corresponds to $\omega,k\rightarrow 0$.
The KPZ equation can be written in the following form 
\begin{equation}
  \label{eq:48}
  F({\bf k},\omega)= G_0({\bf k},\omega) \eta({\bf k},\omega) +
  \frac{1}{2K} G_0({\bf k},\omega)  \int_{\bf q} \int_{\omega}\  {\bf {q}}\cdot 
({\bf {k-q}}) F({\bf {q}},\omega-\Omega)F({\bf {k-q}},\omega),
\end{equation}
which suggests an iterative scheme.  In this equation 
\begin{equation}
  \label{eq:51}
  G_0({\bf
  k},\omega) \equiv \frac{\delta F}{\delta \eta({\bf k},\omega)} =
\frac{1}{{\cal D} k^2-i\omega},
\end{equation}
for the linear equation with ${\cal D}= T/2k$.
The effect of the nonlinear or the second term of the right hand side
of Eq. (\ref{eq:48})  is to change $G_0$ to $G$ (such that $F=G\eta$).
From this $G^{-1}$, the coefficient of $k^2$ in the $k,\omega\rightarrow 0$
limit would give the effective temperature of the
problem.  $G$ may be written as 
\begin{eqnarray}
  \label{eq:50}
  G({\bf k},\omega) &=&G_0({\bf k},\omega) +\frac{2\Delta}{K^2}
  G^2_0({\bf k},\omega)  \int_{\bf q} \int_{\omega}\   \left({\bf
    q}-\frac{{\bf k}}{2}\right)\cdot \left({\bf q} + \frac{{\bf
      k}}{2}\right) \ {\bf
    k}\cdot \left({\bf q} + \frac{{\bf k}}{2}\right)\nonumber\\
 &&\qquad 
  \times G_0({\bf q}-\frac{{\bf k}}{2},\frac{\omega}{2}-\Omega)\, G_0({\bf
    q}+\frac{{\bf k}}{2},\frac{\omega}{2}+\Omega)\, G_0({\bf q}+\frac{{\bf k}}{2},
  -\frac{\omega}{2}-\Omega).
\end{eqnarray}
For $\omega=0$ and $k\rightarrow 0$, the above equation simplifies to
\begin{equation}
  \label{eq:52}
 G^{-1}({\bf k},0)= G^{-1}_0({\bf k},0)\left ( 1- u \frac{d-2}{4d}
   \int dq \ q^{d-3}\right ),
\end{equation}
where
\begin{equation}
  \label{eq:69}
  u=\frac{K\Delta}{T^3}\, L^{2-d},
\end{equation}
is the dimensionless coupling constant in this problem.  Note that this
is the same as in App. \ref{app:rg}.  The power of $q$ in the integral
follows from power counting while the prefactor $(d-2)$ comes from the
angular contributions of the dot products.  In writing this form, the
$d$-dependent solid angle contribution has been absorbed in the
definition of $u$.  The nature of ultraviolet (small distance)
divergence (as $\Lambda\rightarrow\infty$) is apparent in the integral
of Eq. (\ref{eq:52}).  If in the residual integration in Eq.
(\ref{eq:52}), we perform a thin-shell integration between
$\Lambda(1-\delta l)<q<\Lambda$ and set $\Lambda=1$, the effective
temperature for the left-over long wavelength part is given by
\begin{equation}
  \label{eq:53}
  T^{<}=T \left ( 1- u \,\frac{d-2}{4d}\, \delta l \right ).
\end{equation}
On rescaling  $x\rightarrow bx$, we have ${\bf k}\rightarrow (1-\delta l){\bf
k}$ where $b=\exp(\delta l)\approx 1+\delta l$.  The renormalized
temperature is then 
\begin{equation}
  \label{eq:54}
\tilde{T}=b^{(1-2\nu)/\nu} \ T^{<}\approx T^{<}\ \left (1+\delta l
\,  \frac{1-2\nu}{\nu}\right ).   
\end{equation}
Combining Eqs. \ref{eq:54} and \ref{eq:53}, one gets ($\delta l\equiv
\delta L/L$)
\begin{equation}
  \label{eq:55} L\frac{\partial T}{\partial L} = \left
    (\frac{1-2\nu}{\nu} + \frac{2-d}{4d}\, u\right ) T.
\end{equation}
as quoted in Eq. (\ref{eq:1.2.3}).
The renormalization of the disorder strength can be obtained from the
definition
\begin{equation}
  \label{eq:56} 
 \dav{F^*({\bf k},\omega)\,F({\bf k},\omega)}= 2\,
  \tilde\Delta \ G({\bf k},\omega)\,G(-{\bf k},-\omega).
\end{equation}
By using Eqs. \ref{eq:48} and \ref{eq:52} for $F$ and $G$ and
following the same iterative procedure as above,  the flow equation for
$\Delta$ can be determined as
\begin{equation}
  \label{eq:57}
 L\frac{\partial \Delta}{\partial L} = \left (\frac{1-2\theta}{\nu} -d
   + \frac{1}{4} u\right ) \Delta.   
\end{equation}
Combining these two, one gets the flow equation for $u$ 
\begin{equation}
  \label{eq:211}
L\frac{du}{dL} = (2-d)u + \frac{2d-3}{2d} u^2,
\end{equation}
as quoted in  Eq.  (\ref{eq:21}).  The extra $d-2$ factor in the flow
equation for $T$ leads to the stable fixed point at $d=1$.  If the
$d$-dependent $u$ term in the flow of $T$ in Eq. (\ref{eq:55}) is
ignored, the resulting flow equation for $u$ would be similar to Eq.
~(\ref{eq:45}) upto a constant factor for $u$.  As already pointed out
the coefficient of the quadratic term in the flow equation can always
be scaled to $1$ by a redefinition of $u$.  There is a major
difference between the RG of App. \ref{app:rg} and the RG done here.
In App. \ref{app:rg}, $\epsilon=2-d$ is a small parameter and used as
such in various expansions, though ultimately the equations remain
valid for a wider range.  Here however there is no small parameter and
so the approximation cannot be controlled by choosing small
$\epsilon$.

\section{Various flow diagrams}
\label{app:flwec}
We discuss the various possibilities of the flows for an RG flow
equation represented by
\begin{equation}
  \label{eq:74}
  L\frac{du}{dL}=\epsilon \, u + c\, u^2,
\end{equation}
where $u(L)$ is the running coupling constant at length scale $L$.
The equation has two fixed points (fp), $u^*=0, -\epsilon/c$.  
For convenience, let us call the $u=0$ case as the ``free'' problem so
that the $u^*=0$ fp corresponds to the free case.  This fixed point is
to be called the trivial fixed point while the nonzero one is the
nontrivial fp.

\figflw

The behaviour of the coupling constant with length scale 
is  determined by the signs of the two constants ${\epsilon}$ and $c$,
and the initial value $u(L_0)$ at $L=L_0$.

If $\epsilon >0$, then $u$ is a relevant variable at the free fixed
point. while $u$ is irrelevant there if $\epsilon <0$.  Special
situations correspond to $\epsilon=0$ for which $u$ is a marginal
variable. 
  The possibilities we need to consider are
\begin{enumerate}
\item {\bf A:} $\epsilon >0, c>0$.  Here the nontrivial fixed point is
  negative and stable.  There is no fixed point for $u>0$.  A bare
  $u<0$ would then be equivalent to a state described by the
  nontrivial fixed point.  For the random medium problem, $u>0$  and
  so a relevant $u$ flows to large values.  The resulting state cannot
  be described in this approach.   If $u$ represents the interaction
  between two polymers, then $u<0$ in Sec. \ref{sec:rg-flows} or
  App. \ref{app:rg} represents a repulsive interaction while $u>0$ is
  for attraction.  Hence one gets a stable nontrivial phase with a
  repulsive interaction in dimensions $\epsilon>0$ in region A.  For
  one dimension ($d=1$), one may associate this fixed point with a
  fermion (or hard core boson)-like behaviour.  
\item {\bf B:} $\epsilon <0, c>0$.  Change in stability of the fixed
  points. Here the nontrivial fixed point is
  on the positive side and the trivial or the free fixed point is
  stable.  A system with negative $u$ would behave on a long scale
  like a free system and so also for small values of $u>0$.  The
  nontrivial fixed point now represents a critical point so that a
  phase transition can be obtained by tuning $u$.  Large values of $u$
  correspond to a different phase not accessible by this RG flow
  equation because the flow goes to infinity.
\item {\bf C:} $\epsilon <0, c<0$. The situation is similar to Region
  B except that the phase transition is now at a negative value of $u$
  and large negative $u$ phase is not accessible.  All positive $u$
  values are equivalent to the free case (asymptotically free).
\item {\bf D:} $\epsilon >0, c<0$.  Here we see the free system is
  unstable while a positive $u$ case is described by the nontrivial
  fixed point.
\item $\epsilon=0$.   This is the $c$-axis, representing the marginal
  case.  The second order leads to a growth of $u>0$ if $c>0$.  In
  this case,  $u$ is a
  marginally relevant variable while for $c<0$ it is a marginally
  irrelevant variable.  There is no fixed point to describe the
  system.   However, the general trend is that if for $\epsilon=0$,
  there is a marginally relevant variable, then that variable leads to a critical
  behaviour (phase transition) for $\epsilon <0$.
\end{enumerate}

The various possibilities are summarized  in Fig. \ref{fig:flw}.  

In all these cases, if the nontrivial fixed point is stable, then it
represents a ``critical'' phase with characteristic exponents, while
if it is unstable it represents a critical  type phase transition with
its own characteristic exponents.  The reunion behaviour in
Sec. \ref{sec:reunion}  and the reunion exponents\cite{reuni}  are examples of
nontrivial exponents at a stable fixed point.  The
unstable fixed point will be associated with a diverging length scale
with an exponent $\zeta=1/|\epsilon|$ as in Eq. (\ref{eq:6}).

Regions A and C are related by $u\to -u$ but others are distinct.
For the polymer problem, as $d$ (dimensionality) changes, the nature
of flow in the partition function approach (Sec.  \ref{sec:rg-flows}
or App. \ref{app:rg}) goes from region {\bf A} to {\bf B}.  One sees a
new criticality developing for higher $d$ ($d=2-\epsilon$) via a
marginally relevant variable.   The RG flow for the RANI model (Sec. \ref{sec:rani} or
App. \ref{app:rani}) also belongs to this type.    For the KPZ approach (Sec.
~\ref{sec:strong} or App. \ref{app:kpz}) one goes from region {\bf D}
for $d=1$ to region {\bf B}.

\section{On Transfer Matrix}

\label{app:transf}
The directed nature of the directed polymer problem makes it amenable
to a transfer matrix approach.  This is another feather in the cap of
the directed polymer problem.   If we know the partition function at a
point $(x,z)$, we may construct the partition function for the $z+1$th
step because it is completely determined by the information available
at the $z$ step.   The exact form of the transfer matrix would depend
on the particular geometry used.

\figtrf

If we consider the geometry shown in Fig. \ref{fig:1}b, where the $x$
and $z$ axes are along the diagonals of the square lattice, then  the
partition function satisfies (see Fig. ~\ref{fig:trf}a)
\begin{equation}
  \label{eq:70}
  Z(x,z+1)= e^{-\eta(x,z+1)/T}\, [Z(x-1,z) +Z(x+1,z)].
\end{equation}
For the standard geometry (Fig. ~\ref{fig:trf}b) with  $x$ and $z$ axes
along the principal directions of the lattice but if we allow diagonal
steps, then the partition function is given by 
\begin{equation}
  \label{eq:71}
  Z(x,z+1)= e^{-\eta(x,z+1)/T}\, [Z(x,z)+\gamma Z(x-1,z) + \gamma Z(x+1,z)],
\end{equation}
where an extra $\gamma$ factor has been introduced to provide
appropriate elasticity to the polymer.  One may set $\gamma=1$ for a
fully flexible polymer.  The initial condition is 
\begin{equation}
  \label{eq:72}
  Z(x,0)=\delta_{x,0},
\end{equation}
where $\delta_{a,b}$ is the Kronecker delta.
The partition function for a chain of length $N$ starting from $(0,0)$
can be obtained by iterating this equation.  For a given realization
of the randomness $\eta$, the partition function (and therefore any
physical quantity) can be calculated exactly for finite $N$.    For
quenched averaging one has to average over various realizations, and
this is where the exactness of the approach gets lost.

As the length $N$ of the polymer increases, the span of $x$ also
increases linearly so that for $N\rightarrow\infty$ one has to study
an infinitely large matrix.  This allows the possibility of phase
transitions in a seemingly one dimensional problem. For numerical
analysis, special care needs to be taken to keep track of the rapid
growth of the partition function as the length increases.
  
For $T\rightarrow 0$, the problem reduces to determination of the
ground state energy.  One may take the limit $E(x,N)=-\lim_{T\rightarrow
  0} T\ln Z(x,N)$, but a direct approach is also possible.  For
geometry of Fig.  ~\ref{fig:trf}a, the energy can be obtained from
\begin{equation}
  \label{eq:73}
  E(x,z+1)=\min (E(x-1,z)+\eta(x,z+1), E(x+1,z)+\eta(x,z+1)),
\end{equation}
so that the globally minimum energy path is 
\begin{equation}
  \label{eq:75}
  E(N)=\min_{x}  E(x,N).
\end{equation}
Though we are considering the square lattice (in $1+1$ dimensions),
generalization to other lattices and higher dimensions are
straightforward.  Similarly, one may consider cases with random
energies on the bonds instead of vertices.

For the overlap problem of Sec. \ref{sec:over}, the partition function
would satisfy in $1+1$ dimensions
\begin{eqnarray}
  \label{eq:76}
  Z(x_1,x_2,z+1)&=& e^{-(\eta(x_1,z+1)+\eta(x_2,z+1))/T} \times\nonumber\\
&&\qquad \sum_{p=\pm 1}\sum_{q=\pm 1} Z(x_1+p,x_2+q,z) \,
[(1-\delta_{x_1,x_2})+e^{-v/T}\delta_{x_1,x_2}],
\end{eqnarray}
 where the last term involving  $v$ is the Boltzmann factor for the
 interaction on contact ($x_1=x_2$). If $v$ is taken as a random
 quantity chosen from a predetermined distribution, then the same
 transfer matrix can be used to treat the RANI model also.

\section{RG for the RANI model}
\label{app:rani}
We show that the flow equation Eq. (\ref{rani:4}) is exact in the
minimal subtraction scheme using dimensional regularization.  More
details may be found in Ref.  \cite{sm93:1,sm95:2}.

The Hamiltonian needed for $\dav[c]{Z^2}$ is 
\begin{equation}
  \label{eq:59}
  {\cal H}=\frac{1}{2}\int_{0}^{N}dz\sum_{i=1}^{4}
\left (\frac{\partial{\bf{r}}_{i}(z)}{\partial z}\right
)^{^{\scriptstyle 2}}-\bar{r}_0\int_0^N dz\ \delta{\bf (}{\bf r}_{12}(z){\bf
)}\
\delta{\bf (}{\bf r}_{34}(z){\bf )},
\end{equation}
where we have set the effective two body interaction to zero.  For the
cumulant, one need only consider the ``connected'' partition function
for this Hamiltonian.  It is advantageous to consider the Laplace
transform of the $N$-dependent partition function as
\begin{equation}
{\cal Z} = \int_0^{\infty} dN \ e^{-sN} 
\dav[c]{Z^2},\label{eq:lap}
\end{equation}
the Laplace conjugate variable being $s$.

\figapp

An expansion in the coupling constant $\bar{r}_0$ can be arranged like
a ladder (``time-ordered'' diagrams) as shown in Fig. \ref{fig:app}
 The
individual pairs of chains are represented by thick lines.
The horizontal wiggly lines in these diagrams stand for
${\bar{r}}_0$. Such a representation is possible because
the $\delta$ function in $H_2$, Eq. ~(\ref{rani:3}), forces
the members of a pair to have the same ${\bf r}, z$
coordinates.  Each chain is described by the free
distribution (``propagator") $G({\bf r}_f - {\bf r}_i,
z_f - z_i) = [2\pi (z_f - z_i)]^{-d/2}
\exp [ - ({\bf r}_f - {\bf r}_i)^2/2(z_f - z_i) ]$ with end 
points $({\bf r}_f,z_f)$ and $({\bf r}_i, z_i)$.   In conformity with
the current usage we use the word ``propagator'' for the lines.
Two chains are therefore described by
\begin{equation}
G^2 ({\bf r}, z) =(4\pi z)^{-d/2}\ G ({\bf r}, z/2).
\label{eq:gsq}
\end{equation}
This $G^2$ is the propagator for the thick lines.  At each
wiggly line, connecting four chains (all four having the
same chain length $z$), there are two integrations over the
spatial coordinates of the two separate pairs of chains
(thick lines).  The loops formed out of the wiggly lines
are only responsible for the divergence at $d=1$.

In order to trace the algebraic origin of the singularity, note that,
by very nature of the interaction, the spatial integrations associated
with the two thick lines are independent of each other.  Each section
of the thick lines, with $z_1, z_2$ as the end points, in a loop
formed with the wiggly lines, contributes $(z_1 - z_2)^{-d/2}$ from
the identity in Eq. ~(\ref{eq:gsq}).  Since the interaction demands same
$z$ for the two thick lines, the $z$ integrals involve $(z_1 -
z_2)^{-d}$ type factors whose Laplace transform would contribute
$\Gamma(1-d)$ with pole at $d=1$.  The two independent spatial
coordinates which are left out after the successive use of the
normalization $\int d{\bf r} \ G({\bf r}, z) = 1$, lead to a ${\cal
  V}^2$ factor (total volume) for each diagram. The convolution nature
of the $z$ integrals, thanks to the time ordering, leads to a simple
product of the individual Laplace transforms of the integrands,
resulting in a geometric series for ${\cal Z}$.

As an example we consider the two loop diagram of
Fig. \ref{fig:app}(c). After integration over the free ends, we are
left with the following 
\begin{equation}
\bar{r}_0^3
\int_{0}^{N}dz_1\int_{0}^{z_1}dz_2\int_{0}^{z_2} dz_3\, 
\int_{\{r,\ r'\}}G^2({\bf r}_{12}, z_{12}) \,G^2({\bf r}_{23}, z_{23})
\,G^2({\bf r'}_{12}, z_{12}) \,G^2({\bf r'}_{23}, z_{23}).
\end{equation}
Here the subscripts denote the successive points along the thick lines
while the two sets of chains are distinguished by the prime.
In the Laplace space this becomes
\begin{equation}
\bar{r}_0^3\,{\cal{V}}^2\,\Gamma^2(\epsilon')
\,{(4\pi)^{-2d}\,s^{-(2+2\epsilon')}}, \qquad(\epsilon'=1-d).
\end{equation}
This can be generalized to arbitrary orders since only ladder type
diagrams are involved.
 
Defining the dimensionless coupling constant $r_0$ through
an arbitrary length scale $L$ as
$r_0=\bar{r}_0L^{2\epsilon}(4\pi)^{-d}$, $\epsilon=1-d$, we
write the series for ${\cal Z}$ to all orders in $r_0$ as
\begin{equation}
{\cal Z}\!\mid_{\bar{v}_0=0}=(4\pi)^d\, {\cal{V}}^2\, s^{-2} 
\,L^{-2\epsilon'}\, \left[ r_0+ \sum_{n=1}^{\infty}r_0^{n+1}\ (sL^2)^{-n{\epsilon'}}
\,{\Gamma}^n(\epsilon') \right].\label{eq:zf}
\end{equation}
It is clear from the above expression that there is a
divergence at $d=1$ at each order ($>1$). 

A renormalization through minimal subtraction would require absorption
of the poles in $\epsilon'$ through
\begin{equation}
r_0 = r(1 + a_1 r + a_2 r^2 + ...).\label{eq:series}
\end{equation}
with $a_n = \sum_{p=1}^{n} a_{n,p} \epsilon^{-p}$ and $r$
as the renormalized coupling constant. In such a scheme,
$a_{n,p} (p \neq n)$ terms are required to take care of the
subleading divergences.

The geometric series of Eq.  ~(\ref{eq:zf}) guarantees that the removal
of the leading poles is sufficient to remove the subleading ones.
All the divergences can be absorbed by the choice $a_p =
(-\epsilon')^{-p}$ which can be obtained by an explicit order by order
calculation.

The $\beta$ function is therefore exact to all orders in
perturbation series and is given by
\begin{equation}
\beta(r)\equiv L\frac{\partial r}{\partial L}=2 (\epsilon'
r+r^2)\label{eq:br}.
\end{equation}
There are two fixed points: (i) $r=0$ and (ii)
$r^*=-\epsilon'$. 
The bare coupling constant $r_0$ which originates from
$v_0^2\Delta$, where $\Delta$, the variance of the
distribution, is strictly positive, requires a positive
$r$. Therefore, the nontrivial fixed point for $d<1$ in
negative $r$ is unphysical.  It however moves to the
physical domain for $d>1$. 

Exactly at $d=1, \epsilon=0$, $r$ grows with length $L$ as
\begin{equation}
r(L)=r(0) \ \left [1+2 r(0)\ln\frac{L_0}{L}\right
]^{^{-1}},\label{eq:rl}
\end{equation}
$r(0)$ being the coupling at length $L_0$.  Hence, the
disorder is marginally relevant.
For $d>1$, there exists an unstable nontrivial fixed point at
$r = \mid \epsilon \mid$ which separates two distinct
regimes of disorder. If we start with a strong enough
disorder, on the right side of the fixed point, it
increases with length scale, going beyond the perturbative
regime.  This is the strong disorder phase.  On the other
hand, the left side of the fixed point is the weak disorder
regime, since $r$ flows to zero (the stable fixed point).


\end{document}